\documentclass[10pt,conference]{IEEEtran}
\IEEEoverridecommandlockouts
\usepackage{xcolor}
\def\BibTeX{{\rm B\kern-.05em{\sc i\kern-.025em b}\kern-.08em
    T\kern-.1667em\lower.7ex\hbox{E}\kern-.125emX}}

\usepackage{hyperref}
\hypersetup{hidelinks}
\usepackage{algorithmic}
\usepackage{algorithm}
\usepackage{amsmath,amsfonts}
\usepackage{mathtools}
\usepackage{stmaryrd}
\usepackage{algorithmic}
\usepackage{graphicx}
\usepackage{textcomp}
\usepackage{xcolor}
\usepackage{bm}
\usepackage{adjustbox}
\usepackage{tabularx}
\usepackage{float}
\usepackage{colortbl}
\usepackage{booktabs}
\usepackage{multirow}
\usepackage{graphicx}
\usepackage{xspace}
\usepackage[most]{tcolorbox}
\usepackage{balance}
\usepackage[normalem]{ulem}
\usepackage{tabularray}
\usepackage{pgfplots}
\usepackage{threeparttable}
\usepackage{tikz}
\usepackage{enumitem}
\usepackage{pifont}
\usepackage{pgfplots}
\usetikzlibrary{pgfplots.groupplots}
\usepackage{color}
\usepackage{colortbl}
\pgfplotsset{compat=1.18}
\useunder{\uline}{\ul}{}

\newcommand{\greyboxb}[2]{
\vspace{0.05cm}
    \begin{tcolorbox}[
        left=2pt, right=2pt, top=2pt, bottom=2pt,
        boxrule=0.2mm,
        leftrule=2mm,
        arc=0mm,
        colframe=black!40!white,
        colback=black!5!white,
        colbacktitle=black!50!white
    ]
    \textbf{#1}{#2}
    \end{tcolorbox}
}

\newcommand{\phead}[1]{\vspace{1mm} \noindent {\bf #1}}

\newcommand{\toolname}{\textsc{HFuzzer}}
\newcommand{\baseline}{\textsc{GPTFuzzer-A}}

\definecolor{codeborder}{RGB}{240, 240, 240}
\NewTotalTCBox{\code}{ s v }
{verbatim,colback=white,colframe=codeborder,boxsep=0mm,left=1pt,right=1pt,top=1pt,bottom=1pt}
{\lstinline^#2^}
\usepackage{diagbox}
\newcommand{\fix}[1]{{#1}}

\definecolor{darkgreen}{rgb}{0.0, 0.5, 0.0}

\begin{document}
\title{{\toolname}: Testing Large Language Models for Package Hallucinations via Phrase-based Fuzzing} 

\author{\IEEEauthorblockN{Yukai Zhao\IEEEauthorrefmark{1}\IEEEauthorrefmark{2}, Menghan Wu\IEEEauthorrefmark{2}, Xing Hu\thanks{\IEEEauthorrefmark{3}Corresponding Author}\IEEEauthorrefmark{2}\IEEEauthorrefmark{3}, Xin Xia\IEEEauthorrefmark{2}}
\IEEEauthorblockA{
\IEEEauthorrefmark{1}School of Software Technology, Zhejiang University, Ningbo, China \\
\IEEEauthorrefmark{2}The State Key Laboratory of Blockchain and Data Security, Zhejiang University, Hangzhou, China \\
\{yukaizhao2000, menghanwu, xinghu\}@zju.edu.cn, xin.xia@acm.org}
}

\maketitle
\begin{abstract}
Large Language Models (LLMs) are widely used for code generation, but they face critical security risks when applied to practical production due to package hallucinations, in which LLMs recommend non-existent packages.
These hallucinations can be exploited in software supply chain attacks, where malicious attackers exploit them to register harmful packages.
It is critical to test LLMs for package hallucinations to mitigate package hallucinations and defend against potential attacks.
Although researchers have proposed testing frameworks for fact-conflicting hallucinations in natural language generation, there is a lack of research on package hallucinations.
To fill this gap, we propose {\toolname}, a novel phrase-based fuzzing framework to test LLMs for package hallucinations.
{\toolname} adopts fuzzing technology and guides the model to infer a wider range of reasonable information based on phrases, thereby generating enough and diverse coding tasks.
Furthermore, {\toolname} extracts phrases from package information or coding tasks to ensure the relevance of phrases and code, thereby improving the relevance of generated tasks and code.
We evaluate {\toolname} on multiple LLMs and find that it triggers package hallucinations across all selected models.
Compared to the mutational fuzzing framework, {\toolname} identifies \fix{2.60×} more unique hallucinated packages \fix{and generates more diverse tasks.}
Additionally, when testing the model GPT-4o, {\toolname} finds 46 unique hallucinated packages.
Further analysis reveals that for GPT-4o, LLMs exhibit package hallucinations not only during code generation but also when assisting with environment configuration.

\end{abstract}


\section{Introduction}
\label{sec:intro}

Large Language Models (LLMs) have shown significant potential across various domains and are widely used for code generation~\cite{ahmed2025artificial}.
However, despite their success in tackling complex tasks, LLMs face critical challenges related to security and privacy~\cite{yang2024distillseq, siddiq2024quality, siddiq2023sallm,liu2024exploring}.
One major issue is hallucination, where LLM-generated outputs may appear credible or authentic but are factually incorrect, self-contradictory, or unrelated to inputs~\cite{zhang2023siren,gao2025current}.
This issue has been extensively studied in natural language generation (NLG)~\cite{ji2023survey}.
% Recently, Liu et al.~\cite{liu2024exploring} explore it in the context of code generation, categorizing different types of code hallucinations.
Recently, Liu et al.~\cite{liu2024exploring} explore hallucinations in code generation and classify code hallucinations into 19 types.
One critical hallucination type is \textit{package hallucination}, which is defined as LLMs recommend packages or libraries that do not exist~\cite{krishna2025importing,spracklen2024we}.

\begin{figure}[t]
\centerline{\includegraphics[width=0.9\linewidth]{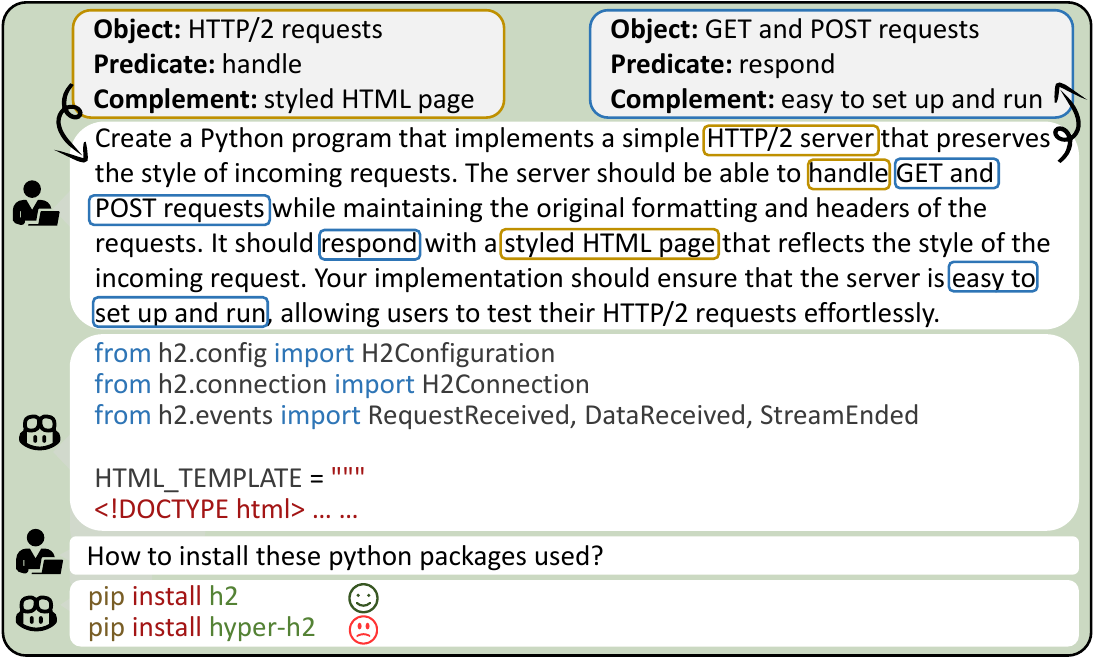}}
\vspace{-0.3cm}
\caption{An Example of Package Hallucination}
\label{Motivation}
\vspace{-0.7cm}
\end{figure}

Compared to other types of code hallucinations, package hallucination poses a higher risk of malicious exploitation, introducing new software supply chain security threats~\cite{packagehallucinations2023,krishna2025importing}.
Such attacks often fall under package obfuscation incidents, where developers are misled into importing packages they do not expect, which is one of the most serious problems in supply chain security~\cite{spracklen2024we, neupane2023beyond}.
% Attackers can exploit LLM package hallucinations by registering malicious packages with the same name as the hallucinated package in the repository.
% Subsequently, other LLM users may be recommended to use these hallucinated packages in their code, leading to security breaches.
Figure~\ref{Motivation} shows an example.
A user first prompts the model to generate a Python program that implements a simple HTTP/2 server with some specific requirements.
After generating the code, the user then asks how to install the packages used in the generated program.
In response, the model recommends two packages to be installed via pip.
Upon inspection, it is found that the package ``h2'' is correct, while ``hyper-h2'' is a hallucinated package that does not exist in the package repository (PyPI~\cite{PYPI}).
If an attacker registers this hallucinated package in the repository and embeds malicious code within it, uninformed developers may inadvertently install and execute it, exposing themselves to supply-chain attacks.
% Attackers could carry out malicious attacks by registering malicious packages with the same names as hallucinated packages in repositories.
% Other LLM users might be misled into using these hallucinated packages in their code based on the model's recommendations.
Moreover, researchers and practitioners have developed various LLM agents for end-to-end software development (such as Devin~\cite{Devin}), which are capable of using tools and executing commands~\cite{wang2024opendevin}. 
This further increases the success rate of package hallucination-based attacks, as malicious packages may be downloaded into the development environment without the developers' awareness.

Spracklen et al.~\cite{spracklen2024we} and Krishna et al.~\cite{krishna2025importing} construct datasets and empirical \fix{studies} on package hallucinations.
While valuable, their studies are limited by the size of the dataset and cannot cover a wide range of code generation scenarios.
Although Drowzee~\cite{li2024drowzee} is proposed to test LLMs for fact-conflicting hallucinations in NLG, research on code hallucinations—particularly on package hallucinations—remains scarce.
Inspired by testing software to help discover failures, we propose to test LLMs for package hallucinations.
However, testing LLMs faces the following two key challenges:

\noindent • \textbf{Challenge~1: How to cover as many code generation scenarios for LLMs as possible?}
To cover these scenarios, we need to generate adequate and diverse coding tasks to test LLMs.
Although Drowzee~\cite{li2024drowzee}, MORTAR~\cite{guo2024mortar}, and MetaQA~\cite{yang2025hallucination} are proposed to generate natural language questions with single correct answers (e.g., ``Did Haruki Murakami and Bob Dylan ever win the same award?''), they cannot complete the generation of coding tasks with multiple correct implementations.
Moreover, approaches that mutate existing tasks using predefined mutation rules often result in limited task diversity, as the variations are bounded by the original task structure and mutation rules. 
Although such strategies may be effective in adversarial attacks like jailbreak attacks, they are inadequate for generating diverse coding tasks.

\noindent • \textbf{Challenge~2: How to generate code-relevant tasks to test LLMs for package hallucinations?}
Non-code-related tasks introduce conflict in the prompt of the model (e.g., asking an LLM to generate code for the task of ``Win the Nobel Prize in literature''), which undermines the effectiveness of testing LLMs for package hallucinations.
Therefore, automatically generated tasks must be restricted to code-related ones.
However, leveraging LLMs or applying simple mutation-based approaches often cannot ensure that the constraint is met, especially when diversity is also desired.
This issue may lead to the generation of non-code-related tasks.

To overcome these challenges, we propose {\toolname}, a novel phrase-based fuzzing framework designed to test LLMs for package hallucinations.
For \textbf{Challenge-1}, {\toolname} adopts fuzzing technology and leverages LLMs as a task generation engine, which enables {\toolname} to generate adequate tasks.
To increase the diversity of tasks, {\toolname} guides LLM to generate tasks based on phrases.
Since LLM has been pre-trained on a large amount of data, when phrases are input, the model can use this knowledge to infer a wider range of reasonable information, thereby improving the diversity of tasks.
For instance, as shown in Figure~\ref{Motivation}, {\toolname} leverages LLM to generate a coding task based on phrases \code{HTTP/2 requests}, \code{handle}, and \code{styled HTML page}.
In the generated task, LLM infers the task ``implement an HTTP/2 server'' based on the phrase \code{HTTP/2 requests}.
% Rather than relying on complete package descriptions, {\toolname} encourages the model to reason from abstract phrases, prompting it to draw on its internal knowledge and generalization capabilities.
% {\toolname} selects three phrases at a time to guide the LLM to generate coding tasks.
% Rather than relying on complete package descriptions, {\toolname} encourages the model to reason from abstract phrases, prompting it to draw on its internal knowledge and generalization capabilities.
% This abstraction promotes greater creativity in task generation.
% To further enhance diversity, {\toolname} employs a power-based selection mechanism: each phrase is assigned a power that reflects its selection probability, and the power values are dynamically updated during execution, which reduces repeated phrase usage.
% For Challenge-2, xxxx.
For \textbf{Challenge-2}, {\toolname} extracts phrases from package information or coding tasks to ensure that the phrases are relevant to the code, thereby improving the relevance of the generated tasks and code.
For instance, the phrase \code{HTTP/2 requests} is more conducive to the model inferring code-related information than \code{Nobel Prize}.
Inspired by the fact that subject-predicate-object triples can summarize information in three phrases, {\toolname} formulates package information or coding tasks as phrase compositions $ \langle Object, Predicate, Complement \rangle$ to obtain richer phrases and avoid redundancy.
Among them, $Object$ represents the object processed by the package or the coding task (e.g., \code{HTTP/2 requests} and \code{Get and Post Requests} in Figure~\ref{Motivation}), $Predicate$ represents the method applied to $Object$ (e.g., \code{handle} and \code{respond} in Figure~\ref{Motivation}), and $Complement$ represents additional relevant details (e.g., \code{styled HTML page} and \code{easy to set up and run} in Figure~\ref{Motivation}).
% To obtain richer phrases and avoid redundancy, inspired by the subject-predicate-object triples, {\toolname} formulates package information or coding tasks as phrase compositions $\langle Object, Predicate, Complement \rangle$, 
% where: 
Based on the extracted phrases, {\toolname} guides LLM to consider the packages related, and then restricts the tasks to those that can be solved by using packages, thus increasing the likelihood of generating code-related tasks that require calling packages.

To evaluate {\toolname}, we use the descriptions of the top 100 Python packages in libraries.io~\cite{libraries} as the initial input and set the budget of a run as 1000 rounds.
\fix{We count the number of unique hallucinated packages to evaluate the effectiveness of {\toolname} and cluster the generated tasks to analyze diversity.}
We compare {\toolname} with {\baseline}, which is adapted from GPTFuzzer~\cite{yu2023gptfuzzer}, and use \fix{nine} models as tester and target models to comprehensively assess the generalizability of {\toolname}.
The tester model is the model used by {\toolname} and {\baseline}; the target model is the model tested.
\fix{Our results} show that {\toolname} successfully triggers package hallucinations in all target models and outperforms {\baseline} across all tester models, \fix{finding on average 2.60x more unique hallucinated packages.
Tasks generated by {\toolname} are more diverse than those from {\baseline}.}
We also find 46 unique hallucinated packages recommended by GPT-4o.
\fix{Further analysis shows that for GPT-4o, package hallucinations occur not only during code generation but also when assisting with environment configuration.}

The contributions of our paper are summarized as follows:
% \vspace{-0.1cm}
\begin{itemize}[leftmargin=*]
    \item We design a new phrase-based coding task generation method that leverages the knowledge of the LLM to infer a wider range of reasonable information based on phrases, thereby generating diverse tasks.
    \item To our knowledge, our framework is the first to introduce the concept of fuzzing into testing LLMs for package hallucinations. The code can be found on our website~\cite{hfuzzer}.
    % to support further studies on package hallucinations.
    \item We conduct a comprehensive evaluation by using different LLMs as tester and target models. Results show that {\toolname} successfully triggers package hallucinations in all target models and outperforms {\baseline} on all tester models. \fix{Tasks generated by {\toolname} are also more diverse than those generated by {\baseline}.}
  %  \item We use {\toolname} to test the model GPT-4o, finding 46 unique hallucinated packages. Our analysis reveals that package hallucinations also occur when the model configures the code execution environment, and are more common than when generating code.
\end{itemize}

\section{Methodology}
\label{sec3}
\begin{figure*}[t]
\centerline{\includegraphics[width=0.77\linewidth]{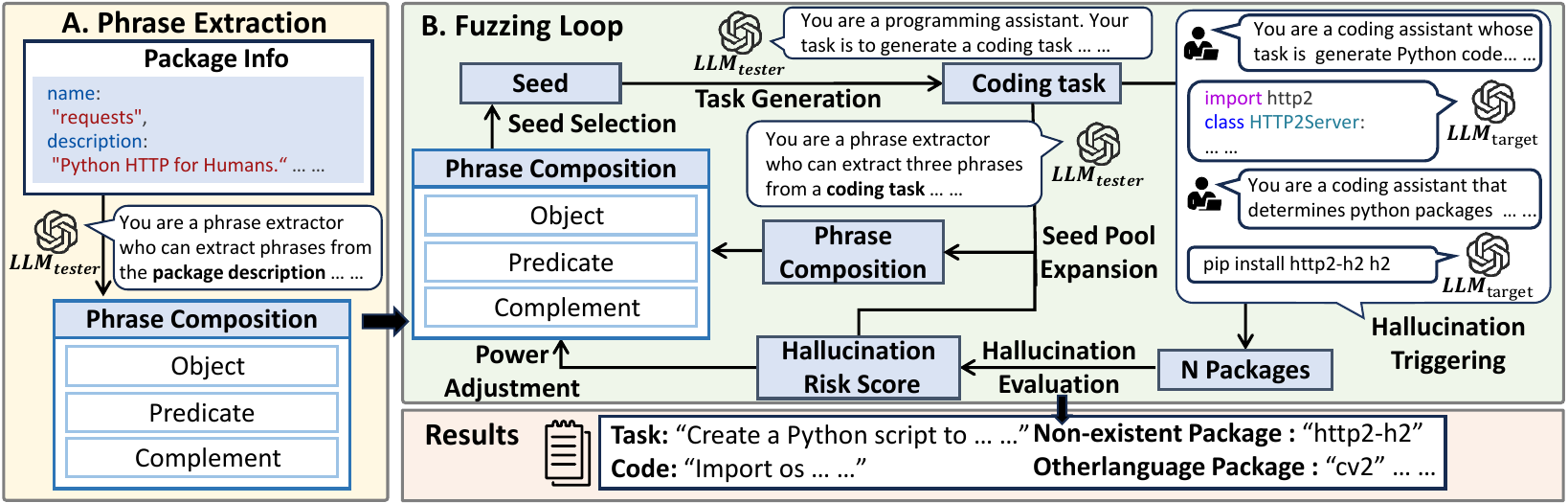}}
\vspace{-0.3cm}
\caption{
The overview of {\toolname}. 
\fix{Phrase Extraction is discussed in Section~\ref{sec:seed_pool},} Seed Selection \fix{is discussed} in Section~\ref{sec:selection}, Task Generation \fix{is discussed} in Section~\ref{sec:generation}, \fix{Hallucination Triggering is discussed} in Section~\ref{sec:query}, \fix{Hallucination Evaluation is discussed} in Section~\ref{sec:check},
Power Adjustment \fix{is discussed in} Section~\ref{sec:modify}, and \fix{Seed Pool Expansion is discussed} in Section~\ref{sec:extract}.
}

% \caption{Overview of {\toolname}}
\label{Overall}
\vspace{-0.7cm}
\end{figure*}

\fix{Figure~\ref{Overall} provides an overview of {\toolname}, which includes two parts (i.e., Phrase Extraction and Fuzzing Loop) and resembles the fuzzing process (Seed Pool Initialization,  Seed Selection, Seed Mutation, and Execution).}
The tester model $LLM_{tester}$ is the model used by {\toolname}, and the target model $LLM_{target}$ is the tested model.
% \fix{The input of {\toolname} is package information, which includes the package names (e.g., ``requests'') and their descriptions (e.g., ``Python HTTP for Humans'').
\fix{{\toolname} uses package information, which includes package names (e.g., ``requests'') and their descriptions (e.g., ``Python HTTP for Humans''), as input and tests closed-source models solely by accessing their input/output.}
\fix{
% {\toolname}'s pipeline resembles the fuzzing process (i.e., Seed Pool Initialization,  Seed Selection, Seed Mutation, and Execution) and solely by accessing the input/output to test closed-source models.
The fuzzer first executes the program under test with inputs from testers to initialize the seed pool (Seed Pool Initialization).
Similarly, in the Phrase Extraction phase, {\toolname} extracts phrase compositions $\langle Object, Predicate, Complement \rangle$ from the package information by $LLM_{tester}$ to construct three corresponding phrase pools, which consist of the seed pool (Section~\ref{sec:seed_pool}).}
Subsequently, {\toolname} enters the Fuzzing Loop phase.
\fix{Similar to selecting seeds from the seed pool (Seed Selection), in each round, {\toolname} selects a seed consisting of three phrases from the seed pool according to the power of phrases (Section~\ref{sec:selection}).}
The power is defined as the potential of a phrase, which determines the probability of the phrase being selected.
\fix{The seed is used by $LLM_{tester}$ to generate a coding task (Section~\ref{sec:generation}).
Unlike traditional fuzzing generates new inputs by mutating seeds (Seed Mutation), {\toolname} generates new coding tasks by recombining different phrases. 
In the Hallucination Triggering phase, {\toolname} asks $LLM_{target}$  to generate code for the provided task, and then guides it to recommend packages required to execute code, which corresponds to the Execution of fuzzing.}
{\toolname} performs hallucination evaluation on packages recommended by $LLM_{target}$ to calculate the Hallucination Score (HS), which is used to measure the package hallucination triggering on $LLM_{target}$ (Section~\ref{sec:check}), and adjusts the power of related phrases according to the HS (Section~\ref{sec:modify}), thereby guiding future selection toward under-explored phrases.
To expand the seed pool, for tasks that trigger package hallucinations, {\toolname} extracts new phrase compositions from them and adds new phrase compositions to the seed pool (Section~\ref{sec:extract}).
The process continues until the query budget is exhausted.
Finally, we get the generated coding tasks and corresponding model outputs.
% \fix{In the above process, {\toolname} discards intermediate results if the output format is invalid (e.g., missing tags) to handle incorrect responses, and verifies packages by querying the package index (e.g., PyPI) to avoid false positives due to $LLM_{tester}$ output errors.
% $LLM_{tester}$ is also allowed to respond with ``None'' when unable to generate reasonable output, and $LLM_{target}$ is allowed to refuse unrelated coding tasks.
% Additionally, compared to traditional parsing methods, LLMs provide stronger semantic understanding and natural language processing, allowing them to handle complex inputs.}
\fix{In the above process, to handle incorrect responses, {\toolname} discards intermediate results if the output format is invalid (e.g., missing tags) and allows $LLM_{tester}$ to respond with "None" when it is unable to generate reasonable output. {\toolname} verifies packages by querying package indices (e.g., PyPI) to avoid false positives due to $LLM_{tester}$ output errors. $LLM_{target}$ is also allowed to refuse unrelated coding tasks. Additionally, compared to traditional parsing methods, LLMs provide stronger semantic understanding and natural language processing, allowing them to handle complex inputs.}

\subsection{Phrase Extraction}
\label{sec:seed_pool}
{\toolname} formulates package information as the phrase composition $\langle Object, Predicate, Complement \rangle$ and extracts the phrase composition from each package’s information to construct the seed pool.
This seed pool is divided into three corresponding phrase pools with power, each containing phrases of one component type.
The three phrases of the composition are defined as follows:

\noindent • $Object$ represents the object processed by the package.

\noindent • $Predicate$ represents the method applied by the package.

% \noindent • $Complement$ refers to additional relevant details used to ensure that important information is not overlooked.
\noindent • \fix{$Complement$ represents a phrase that captures essential contextual details beyond the $Object$ or $Predicate$, providing additional context when applicable.}

To achieve this, {\toolname} leverages LLM to extract phrase compositions from package information.
{\toolname} configures the system prompt to define the target of phrase extraction from package information and provides an example to specify the expected response format.
In the user prompt, {\toolname} provides the package information to the model to obtain phrases.
These phrases are added to their respective phrase pools and assigned an initial power, influencing the selection process described in Section~\ref{sec:selection}.
% Note that to handle the case where the model cannot extract a specific phrase when extracting phrases, we prompt the model to replace it with None when the corresponding phrase cannot be extracted.
To handle cases where the model cannot extract a specific phrase due to insufficient information, the prompt instructs the model to answer ``None''.
This design helps minimize the negative impact of incomplete package descriptions on phrase quality.

\noindent \textbf{Example A.} The description of package \code{pre-commit} is ``\textit{A framework for managing and maintaining multi-language pre-commit hooks}''.
We formalize it as $\langle$\textit{pre-commit hooks, managing and maintaining, multi-language support}$\rangle$. Among them, \code{pre-commit hooks} represents the objects processed by this package,  \code{managing and maintaining} represents the methods provided by this package, and  \code{multi-language support} supplements the features of this package.
\fix{{\toolname} extracts such information from package information to construct the seed pool.}
\subsection{Fuzzing Loop}
\label{sec:loop}
Similar to fuzzing, we consider the following process as one round and repeat it until the budget is exhausted.

% 
% \label{sec:codetask}
\subsubsection{Seed Selection}
\label{sec:selection}
In directed fuzzing, fuzzers assign scores to the seeds and prioritize those with higher scores for mutation.
{\toolname} adopts a similar strategy for seed selection.
In this paper, each seed consists of three phrases corresponding to the phrase composition described in Section~\ref{sec:seed_pool}, to increase the amount of information and provide richer context for LLM.
% In seed selection, {\toolname} follows a similar concept, maintaining three phrase pools with power as the seed pool.
%, maintaining a seed pool with the power. 
% Each seed comprises three phrases, corresponding to the phrase composition in Section~\ref{sec:seed_pool}.
Specifically, {\toolname} applies a weighted random selection algorithm to construct a seed: one phrase is selected from each phrase pool, where the selection probability of a phrase is proportional to its power relative to the total power of the phrase pool.
% Specifically, for each phrase pool, the selection probability of a phrase is equal to the ratio of its power to the total power of all phrases within this pool.
The power is initialized in Section~\ref{sec:seed_pool} and continuously adjusted to reduce the likelihood of repeatedly selecting the same phrases.
The details of power adjustment are presented in Section~\ref{sec:modify}.

\noindent \textbf{\fix{Example B1.}}
\fix{In Figure~\ref{Motivation}, {\toolname} selects one phrase from each of three phrase pools to construct a phrase composition: $\langle HTTP/2~requests,~handle,~styled~HTML~page\rangle$.}

% As shown in Figure~\ref{Motivation}, {\toolname} selects a phrase from each of the three phrase pools to form a new phrase composition: $\langle HTTP/2~requests,~handle,$ $styled~HTML~page\rangle$.}

\subsubsection{Task Generation}
\label{sec:generation}
{\toolname} exploits LLM to generate a coding task based on a selected seed.
Specifically, {\toolname} leverages LLM to identify packages that may be relevant to the provided phrases and to generate a coding task that can be solved using those packages.
To enable automatic extraction of the task from the model’s response, {\toolname} includes a formatting example in the prompt and requires the LLM to enclose the generated task within a predefined tag.
During execution, {\toolname} inserts three phrases of the selected seed into a preset user prompt, queries the model, and parses the tagged output to extract the final coding task.

\noindent \textbf{\fix{Example B2.}}
\fix{Based on the phrase composition, {\toolname} uses $LLM_{tester}$ to generate a coding task, shown in Figure~\ref{Motivation}.
% This task aims to implement a simple HTTP/2 server (}\code{HTTP/2requests}\fix{) that can handle GET and POST requests (}\code{handle}\fix{) and respond with a styled HTML page (}\code{style HTML page}\fix{).
This task requires implementing an HTTP/2 server (HTTP/2 requests) that can handle GET and POST requests (handle) and respond with a styled HTML page (style HTML page).}

% \subsubsection{Query}
% % toolname模拟大模型使用场景，触发目标模型的包幻觉。
% % 
% {\toolname} 
% To evaluate whether a coding task can trigger package hallucinations, we design a code agent that simulates LLM usage scenarios.
% The agent takes a coding task as input, and through multiple times of queries, finally outputs packages recommended by the target model.
% Specifically, the agent first guides the target model to generate code that solves the given task, providing an example to clarify the expected response format.
% To mitigate the influence of unintended coding tasks (e.g., non-code-related tasks), the prompt explicitly instructs the model to return ``None'' if it cannot produce Python code.
% It then extracts the generated code snippet using regular expressions and issues a follow-up query to the model, asking for installation commands corresponding to the required packages. The returned command is then parsed to extract packages.
% This process mirrors real-world usage, where developers typically do not directly ask LLM to recommend packages.
% Instead, they first generate code for a task and then query guidance on installing missing packages.

% new
\subsubsection{Hallucination Triggering}

To trigger package hallucination in the common LLM usage scenario, we divide the entire triggering process into two stages: Code Generation and Package Recommendation.
In Code Generation, {\toolname} guides the target model to generate code that solves the generated task, providing an example to clarify the expected response format.
To mitigate the influence of unintended coding tasks (e.g., non-code-related tasks), the prompt explicitly instructs the model to return ``None'' if it cannot produce code.
In Package Recommendation, {\toolname} prompts the target model with the generated code and requires the model to answer the installation commands corresponding to the packages used in the code.
Based on the package answered, {\toolname} further performs hallucination evaluation.
\fix{This design simulates a realistic user scenario (developer requests a code snippet along with the required packages) and reduces false positives from aliases and similar ambiguities compared with rule-based package extraction.
% Additionally, Package Recommendation tests LLMs on the environment configuration.
% Existing research has used LLM to configure the environment (e.g., Docker builds~\cite{hu2025llm}).
Additionally, Package Recommendation tests LLMs on environment configuration, motivated by existing studies that use LLMs for this purpose (e.g., in automatic Docker builds [20]).
It also supports analyses of hallucinations from mismatches between imported modules (e.g., cv2) and actual package names (e.g., opencv-python)~\cite{spracklen2024we}.}

\noindent \textbf{\fix{Example B3.}}
\fix{In Figure~\ref{Motivation}, {\toolname} first prompts $LLM_{target}$ to generate a Python program for the task, and then queries it to return the corresponding package installation commands (i.e., pip install h2 and pip install hyper-h2).}

\label{sec:query}

\subsubsection{Hallucination Evaluation}
\label{sec:check}

In this phase, {\toolname} evaluates the package hallucination based on extracted packages in Section~\ref{sec:query} and calculates the HS.
To achieve fine-grained evaluation, we first classify the extracted packages.
Prior studies define hallucinated packages as those recommended by LLMs but either (i) registered after the model’s knowledge cutoff date or (ii) not present in the appropriate package repository~\cite{krishna2025importing}.
\fix{Since {\toolname} aims not to test whether LLMs can infer unknown packages but rather to find hallucinations within existing knowledge, it adopts the above definition and further classifies hallucinated packages into two types: nonexistent packages and otherlanguage packages.}
% Considering that the goal of {\toolname} is not to test whether LLMs can infer unknown packages, but to find hallucinations within existing knowledge, {\toolname} follows the above definition and further classifies hallucinated packages into two types: non-existent packages and other-language packages.}
% According to this definition, {\toolname} further classifies hallucinated packages into two types: non-existent packages and other-language packages. 
Additionally, we observe a common scenario in which LLMs mistakenly treat standard libraries (e.g., json) as packages requiring installation.
Although these libraries are not found in repositories, they do not strictly qualify as hallucinated packages.
Therefore, we exclude them from hallucinated packages.
In summary, we classify the packages recommended by the model into four types: ``stdPackage'', ``existPackage'', ``otherLanguagePackage'', and ``nonExistentPackage''.
To formalize these types, let $ P_{std} $ be the set of standard libraries, $ P_{exist} $ be the set of packages registered in the appropriate package repository before the model's knowledge cutoff date, and $ P_{\text{lib}} $ be the set of packages in Libraries.io~\cite{libraries}.
We define these types as follows:

\begin{itemize}
    \item Package $ p $ is classified as a ``stdPackage'' if $p \in P_{\text{std}}.$

    \item Package $p$ is classified as an ``existPackage'' if $p \in P_{\text{exist}}.$
    \item Package $p$ is classified as an ``otherLanguagePackage'' if
    $
    p \notin (P_{\text{std}} \cup P_{\text{exist}}) \land p \in P_{\text{lib}}.
    $

    \item Package $ p $ is classified as a ``nonExistentPackage'' if
    $
    p \notin (P_{\text{std}} \cup P_{\text{exist}} \cup P_{\text{lib}}).
    $
\end{itemize}
% \begin{itemize}
%     \item A package $ p $ is classified as a ``stdPackage'' if
%     \[
%     p \in P_{\text{std}}.
%     \]

%     \item A package $p$ is classified as an ``existPackage'' if
%     \[
%     p \in P_{\text{exist}}.
%     \]

%     \item A package $p$ is classified as an ``otherLanguagePackage'' if
%     \[
%     p \notin (P_{\text{std}} \cup P_{\text{exist}}) \land p \in P_{\text{lib}}.
%     \]

%     \item A package $ p $ is classified as a ``nonExistentPackage'' if
%     \[
%     p \notin (P_{\text{std}} \cup P_{\text{exist}} \cup P_{\text{lib}}).
%     \]
% \end{itemize}
\noindent \fix{Libraries.io is a cross-language package index that aggregates repositories from multiple programming ecosystems.
Following prior studies~\cite{spracklen2024we,krishna2025importing}, we use it to verify whether a package originates from a different language.}

Based on the above definition, {\toolname} classifies packages recommended by the model and calculates the HS to measure the package hallucination triggering on the target model.
Let $ N_{\text{package}}$ be the total number of packages recommended by the target model, $N_{\text{non}}$ be the number of packages classified as ``nonExistentPackage'', and $N_{\text{other}}$ be the number of packages classified as ``otherLanguagePackage''.
The HS is defined as:
\begin{equation}
\label{eq1}
% \vspace{-0.1cm}
HS = \frac{\alpha \cdot N_{\text{non}}}{N_{\text{package}}}+\frac{\beta \cdot N_{\text{other}}}{N_{\text{package}}} 
\end{equation}
where $\alpha = 1$ and $\beta = 0.5$\footnote{All parameters are determined through an initial experiment.}.
The impact of using nonexistent packages is more serious because it means that the model constructs packages that do not exist, whereas using otherlanguage packages may be due to the model confusing the language of the package.
Thus, {\toolname} sets a larger constant to $\alpha$.

\noindent \textbf{\fix{Example B4.}}
\fix{In Figure~\ref{Motivation}, $LLM_{target}$ recommends two packages, which are classified as an ``existPackage'' and a ``nonExistentPackage''. Hence, the HS is 0.5.}

\subsubsection{Power Adjustment}
\label{sec:modify}

To increase the diversity of coding tasks, {\toolname} adjusts the power of phrases based on the HS.
Let $Power_{o}$ be the original power of the phrase, $N_{\text{new}} $ be the number of hallucinated packages found for the first time in the fuzzing loop, and $N_{\text{old}} $ be the number of hallucinated packages found in the previous rounds.
The adjusted power of the phrase $Power$ is defined as follows:
\begin{equation}
\label{eq3}
\resizebox{0.85\linewidth}{!}{
$
Power = 
\begin{cases}
 Power_{o}\cdot \left (  \frac{k_{1} \cdot HS\cdot N_{\text{new}} }{N_{\text{old}}+N_{\text{new}}}+k_2 \right ) &,~\text{if } N_{\text{old}}+N_{\text{new}} >  0 \\
 Power_{o}\cdot k_2 &,~\text{elif}~N_{\text{package}} > 0\\
  Power_{o}\cdot k_3 &,~\text{otherwise}
\end{cases}
$
}
\end{equation}
where $k_1 = 0.15$, $k_2 = 0.8$, and $k_3 = 0.6$.
{\toolname} reduces the power of phrases corresponding to the task that do not recommend packages by a factor $k_3$, while the power of phrases that do not recommend hallucinated packages is decreased by a factor $k_2$, thereby lowering their chance of reselection. 
For phrases where the task triggers package hallucination, the power reduction is determined by the HS and the ratio $\frac{N_{\text{new}}}{N_{\text{old}}+N_{\text{new}}}$.
By incorporating HS, {\toolname} fine-tunes power based on the specific response of the target model.
{\toolname} also assigns relatively higher power to phrases that find new hallucinated packages by $\frac{N_{\text{new}} }{N_{\text{old}}+N_{\text{new}}}$.
Intuitively, similar to seeds for generating new coverage in fuzzing, these phrases are more likely to generate coding tasks that can trigger package hallucination in the model.

\noindent \textbf{\fix{Example B5.}}
\fix{Based on the HS (0.5), assuming the old power is 1 and the hallucinated package is the first appearance, the updated power is 0.875.}
% For phrases where the task triggers package hallucination, the power reduction is determined by the ratios $\frac{N_{non}}{N_{p}}$, $\frac{N_{other}}{N_{p}}$, and $\frac{N_{new}}{N_{non}+N_{other}}$. 

% Additionally, {\toolname} assigns relatively higher power to phrases that find new hallucinated packages, as indicated by $\frac{N_{new} }{N_{non}+N_{other}}$.
% Therefore, similar to the emphasis placed on seeds that produce new coverage in fuzzing, {\toolname} gives relatively higher power to phrases that find new hallucinated packages, and rewards phrases that find more non-existent packages.
% Our intuition is that similar to seeds for generating new coverage in fuzzing, these phrases are more likely to generate coding tasks that can trigger package hallucination in the model.

% The specific impact on the results is determined by the parameters $\alpha$ and $\beta$.
% We believe that the impact of using a non-existent package is more serious because it means that the model constructs a package name that does not exist, while using other-language packages may be due to the model confusing the language of the package.
% Based on the above view, in the parameter setting, {\toolname} uses a larger parameter for $\alpha$.
% By incorporating these ratios, {\toolname} fine-tunes power based on the types of hallucinated packages.

\subsubsection{Seed Pool Expansion}
\label{sec:extract}
Traditional fuzzers generate more inputs through mutation.
In contrast, {\toolname} leverages LLMs to generate new coding tasks based on phrase compositions.
% However, the phrase combinations in the initial seed pool are limited.
However, the phrases in the initial seed pool come from the input of Section~\ref{sec:seed_pool}, and their compositions are limited.
As the number of running rounds increases, the probability of repeated compositions will also increase, thereby reducing the diversity of generated tasks.
% However, relying solely on the initial seed pool limits the scalability of {\toolname} due to a limited set of possible combinations.
To overcome this, {\toolname} instructs LLM to extract phrase compositions from coding tasks.
% and enabling the generation of an unlimited variety of distinct tasks.
% Specifically, {\toolname} instructs LLM to extract phrase compositions from coding tasks.
% and  provides the model with phrases used to generate the corresponding coding task and instructs the model to extract as many distinct phrases as possible.
% 另外，和Section3.1一样，{\toolname}为新添加的短语赋予了初始权重。
% 这个权重受同样受Package分类结果影响。
Additionally, as in Section~\ref{sec:seed_pool}, {\toolname} assigns initial power to newly extracted phrases.
% This weight is also affected by the Package classification results.
Let $Power_{\text{initial}}$ be the initial power used in Section~\ref{sec:seed_pool}.
The power of newly extracted phrases $Power_{\text{new}}$ is defined as follow:
\begin{equation}
\vspace{-0.1cm}
\label{eq4}
\resizebox{0.75\linewidth}{!}{
$
Power_{\text{new}} = Power_{\text{initial}}\cdot \left ( k +  \frac{(1 - k) \cdot N_{\text{new}}}{N_{\text{new}}+N_{\text{old}}} \right )
$
}
% \vspace{-0.1cm}
\end{equation}
where $k=0.6$.
Unlike $Power$ in Formula~\ref{eq3}, $Power_{\text{new}}$ places more emphasis on finding new hallucinated packages.
Therefore, we enlarge the impact of the proportion of using new hallucination packages on power. After preliminary experiments, we set  $k$ as 0.6.

\noindent \textbf{\fix{Example B6.}}
\fix{In Figure~\ref{Motivation}, {\toolname} extracts new phrase compositions (i.e., GET and POST requests, respond, and easy to set up and run) from the task. Their power is 1.0.}

At the end of each round, {\toolname} records the generated coding task, the output of the target model, and the hallucination evaluation results, which can be used to reproduce the package hallucinations found during the test.

% As shown in Figure~\ref{Motivation}, {\toolname} extracts new phrase compositions (i.e., }\code{GET and POST requests}\fix{, }\code{respond}\fix{, and} \code{easy to set up and run}\fix{) from the tasks. All of them are assigned a power of 1.0.

\section{Evaluation}
\label{evaluation}
In this section, we study the following research questions:

% \phead{RQ1: How effective are the tasks generated by {\toolname} for triggering LLM's package hallucinations?} This RQ studies the effectiveness of the coding tasks generated by {\toolname} in triggering package hallucinations.

% \phead{RQ2: How effective is {\toolname} in testing LLMs for package hallucinations?} This RQ studies whether {\toolname} is more effective in testing LLMs for package hallucinations compared to the baseline.
\phead{\fix{RQ1: How effective is {\toolname} in testing LLMs for package hallucinations?}} \fix{This RQ studies {\toolname}'s ability to trigger package hallucination on different models, and evaluates whether it is more effective in testing LLMs for package hallucination compared with {\baseline}.}

\phead{\fix{RQ2: Whether the tasks generated by {\toolname} more diverse?}} \fix{This RQ studies the diversity of tasks generated by {\toolname} and {\baseline}.}

\phead{RQ3: Whether each module of {\toolname} contributes to its performance?} This RQ explores the impact of different modules of {\toolname} on its performance.

\subsection{Experimental Setup}
\noindent\textbf{Implementation.}
In our implementation, we access the model through the API provided by the OpenAI library and obtain package information through the APIs of the package repository and libraries.io~\cite{libraries}.

\noindent \textbf{Baseline.}
To our knowledge, no study has focused on testing LLMs for package hallucinations. 
Existing studies mainly concentrate on empirical studies and mitigating related hallucinations.
Note that {\toolname} is intended to test LLMs for hallucinations, not to detect whether the model output contains hallucinations.
The study closest to ours is Drowzee~\cite{li2024drowzee}, which tests LLMs for fact-conflicting hallucinations in NLG through metamorphic testing.
However, it assumes a unique answer and cannot be applied to package hallucinations. 
Therefore, we compare {\toolname} with {\baseline}, which is adapted from GPTFuzzer~\cite{yu2023gptfuzzer}.
\fix{We modify the mutation operators and the initial seeds of GPTFuzzer, and choose a random strategy for seed selection.
The mutation operators are adapted to generate new coding tasks instead of new templates. The initial seeds, originally jailbreak templates, are replaced with coding tasks derived from package descriptions to ensure input consistency between {\toolname} and {\baseline}.}

\noindent \textbf{Language Selection.}
We choose Python as the target language to evaluate {\toolname}.
Python is a popular language, ranked number one on the TIOBE index~\cite{tiobe}, with a well-developed ecosystem and widespread use in related studies~\cite{krishna2025importing,spracklen2024we,liu2024exploring}.

\noindent \textbf{Model Selection.}
To comprehensively evaluate {\toolname} on LLMs with different training data and architectures, we select multiple open-source popular models \fix{and closed-source models.}
The detailed information is shown in Table~\ref{models}.

% check
\begin{table}[h]
\vspace{-0.4cm}
\caption{The Details of Selected Models}
\vspace{-0.2cm}
\label{models}
\centering % 添加这个命令使表格居中
\resizebox{0.9\linewidth}{!}{
\begin{tabular}{@{}c|c|c|c|c@{}}
\toprule
\rowcolor[HTML]{FFFFFF} 
\multicolumn{1}{c|}{\cellcolor[HTML]{FFFFFF}\textbf{Model}} & \multicolumn{1}{c|}{\cellcolor[HTML]{FFFFFF}\textbf{Size}} & \multicolumn{1}{c|}{\cellcolor[HTML]{FFFFFF}\textbf{Code Model}} & \multicolumn{1}{c|}{\cellcolor[HTML]{FFFFFF}\textbf{Open Weights}} & \multicolumn{1}{c}{\cellcolor[HTML]{FFFFFF}\textbf{Full Name}} \\ \midrule
\rowcolor[HTML]{EFEFEF} 
Meta-Llama-3                                       & 8B                                                  & No                                                       & Yes                                                        & Meta-Llama-3-8B-Instruct~\cite{llama3}                              \\
\rowcolor[HTML]{FFFFFF} 
Qwen2.5-Coder                                      & 7B                                                  & Yes                                                      & Yes                                                        & Qwen2.5-Coder-7B-Instruct~\cite{hui2024qwen2}                             \\
\rowcolor[HTML]{EFEFEF} 
DeepSeek-Coder                                    & 6.7B                                                & Yes                                                      & Yes                                                        & DeepSeek-Coder-Instruct 6.7B~\cite{guo2024deepseek}                          \\
\rowcolor[HTML]{FFFFFF} 
Meta-Llama-3.1                                     & 8B                                                  & No                                                       & Yes                                                        & Meta-Llama-3.1-8B-Instruct~\cite{llama3.1}                          \\
\rowcolor[HTML]{EFEFEF} 
Mistral-v0.3                                            & 7B                                                  & No                                                       & Yes                                                        & Mistral-7B-Instruct-v0.3~\cite{jiang2023mistral}                              \\
\rowcolor[HTML]{FFFFFF} 
\fix{Meta-Llama-3.3}                                       & \fix{70B}                                           & \fix{No}                                            &\fix{Yes}                                           & \fix{Meta-Llama-3.3-70B-Instruct~\cite{llama3.3}}                                          \\
\rowcolor[HTML]{EFEFEF} 
\fix{DeepSeek-V3}                                   & \fix{671B}                                   & \fix{No}                              & \fix{Yes}                                 &\fix{DeepSeek-V3~\cite{liu2024deepseek}}                           \\ 
\rowcolor[HTML]{FFFFFF} 
GPT-4o mini   & -            & No    & No  & GPT-4o mini~\cite{achiam2023gpt}          \\
\rowcolor[HTML]{EFEFEF} 
\fix{GPT-4o}  & \fix{-}  & \fix{No}  & \fix{No}  & \fix{GPT-4o~\cite{hurst2024gpt}}                                          \\

\bottomrule
\end{tabular}
}
\vspace{-0.2cm}
\end{table}
\begin{table*}[!ht]
\centering
\setlength{\abovecaptionskip}{0.05cm} %# 调整间距
\setlength{\belowcaptionskip}{0.1cm}
\renewcommand{\arraystretch}{1.1}
\setlength{\tabcolsep}{2pt}
\setlength{\belowcaptionskip}{0.1cm}
\caption{RQ1: Unique Hallucinated Packages Results}
\vspace{-0.1cm}
\resizebox{0.9\textwidth}{!}{
\begin{tabular}{cc|cl|cl|cl|cl|cl|cl|cl|cl|cl|c}
\hline
\multicolumn{2}{c|}{\multirow{2}{*}{\diagbox[width=11em,trim=l]{Tester}{Target}}} & \multicolumn{2}{c|}{Meta-Llama-3} & \multicolumn{2}{c|}{Qwen2.5-Coder}  & \multicolumn{2}{c|}{DeepSeek-Coder}& \multicolumn{2}{c|}{Meta-Llama-3.1} & \multicolumn{2}{c|}{Mistral-v0.3}  & \multicolumn{2}{c|}{\fix{Meta-Llama-3.3}} & \multicolumn{2}{c|}{\fix{DeepSeek-V3}} & \multicolumn{2}{c|}{GPT-4o mini}  & \multicolumn{2}{c|}{\fix{GPT-4o}} & \multirow{2}{*}{\begin{tabular}[c]{@{}c@{}}Avg. \\ (same Tester)\end{tabular}} \\ \cline{3-20}
\multicolumn{2}{c|}{} & $P_{uniq}$   & $R_{inc}$   & $P_{uniq}$ & $R_{inc}$ & $P_{uniq}$ & $R_{inc}$  & $P_{uniq}$    & $R_{inc}$    & $P_{uniq}$    &$R_{inc}$   &\fix{$P_{uniq}$} & \fix{$R_{inc}$}& \fix{$P_{uniq}$}  &\fix{$R_{inc}$}  & $P_{uniq}$     &$R_{inc}$& \fix{$P_{uniq}$} &\fix{$R_{inc}$} &  \\ \hline
\multicolumn{1}{c|}{}  & \fix{G-A}   & 23   &   & 5  & & 14  &     & 44&  & 54&  & \fix{3}  &  & \fix{0}&     & 5  &     & \fix{0} &     &     \\
\multicolumn{1}{c|}{\multirow{-2}{*}{Meta-Llama-3}}    & \fix{H} & 51 & \multirow{-2}{*}{\tikz \draw[red, thick, ->] (0,0) -- (0,0.2);~2.22} & 6 & \multirow{-2}{*}{\tikz \draw[red, thick, ->] (0,0) -- (0,0.2);~1.20} & 24 & \multirow{-2}{*}{\tikz \draw[red, thick, ->] (0,0) -- (0,0.2);~1.71} & 65& \multirow{-2}{*}{\tikz \draw[red, thick, ->] (0,0) -- (0,0.2);~1.48} & 95& \multirow{-2}{*}{\tikz \draw[red, thick, ->] (0,0) -- (0,0.2);~1.76} & \fix{6}  & \multirow{-2}{*}{\tikz \draw[red, thick, ->] (0,0) -- (0,0.2);~\fix{2.00}} &\fix{3} & \multirow{-2}{*}{\tikz \draw[red, thick, ->] (0,0) -- (0,0.2);~~~\fix{-}} & 7  & \multirow{-2}{*}{\tikz \draw[red, thick, ->] (0,0) -- (0,0.2);~1.40}    &\fix{1} & \multirow{-2}{*}{\tikz \draw[red, thick, ->] (0,0) -- (0,0.2);~~~\fix{-}} & \multirow{-2}{*}{\fix{1.68}}\\ \hline
\multicolumn{1}{c|}{}    & \fix{G-A}& 29 &  & 1 &  & 19 &  & 47&  & 55&  & \fix{2}  &  & \fix{1} &     & 1  &     & \fix{0} &     &     \\
\multicolumn{1}{c|}{\multirow{-2}{*}{Qwen2.5-Coder}}   & \fix{H} & 48 & \multirow{-2}{*}{\tikz \draw[red, thick, ->] (0,0) -- (0,0.2);~1.66} & 6 & \multirow{-2}{*}{\tikz \draw[red, thick, ->] (0,0) -- (0,0.2);~6.00} & 40 & \multirow{-2}{*}{\tikz \draw[red, thick, ->] (0,0) -- (0,0.2);~2.11} & 72& \multirow{-2}{*}{\tikz \draw[red, thick, ->] (0,0) -- (0,0.2);~1.53} & 86& \multirow{-2}{*}{\tikz \draw[red, thick, ->] (0,0) -- (0,0.2);~1.56} & \fix{6}  & \multirow{-2}{*}{\tikz \draw[red, thick, ->] (0,0) -- (0,0.2);~\fix{3.00}} & \fix{3}& \multirow{-2}{*}{\tikz \draw[red, thick, ->] (0,0) -- (0,0.2);~\fix{3.00}}    & 4  & \multirow{-2}{*}{\tikz \draw[red, thick, ->] (0,0) -- (0,0.2);~4.00}    & \fix{0} & \multirow{-2}{*}{~~~\fix{-}}   & \multirow{-2}{*}{\fix{2.65}}\\ \hline
\multicolumn{1}{c|}{}    & \fix{G-A}& 13 &  & 1 &  & 30 &  & 27&  & 40&  & \fix{2}  &  & \fix{0} &     & 0  &     & \fix{0} &     &     \\
\multicolumn{1}{c|}{\multirow{-2}{*}{DeepSeek-Coder}}  & \fix{H} & 36 & \multirow{-2}{*}{\tikz \draw[red, thick, ->] (0,0) -- (0,0.2);~2.77} & 8 & \multirow{-2}{*}{\tikz \draw[red, thick, ->] (0,0) -- (0,0.2);~8.00} & 111& \multirow{-2}{*}{\tikz \draw[red, thick, ->] (0,0) -- (0,0.2);~3.70} & 114     & \multirow{-2}{*}{\tikz \draw[red, thick, ->] (0,0) -- (0,0.2);~4.22} & 184     & \multirow{-2}{*}{\tikz \draw[red, thick, ->] (0,0) -- (0,0.2);~4.60} & \fix{15}& \multirow{-2}{*}{\tikz \draw[red, thick, ->] (0,0) -- (0,0.2);~\fix{7.50}} & \fix{3} & \multirow{-2}{*}{\tikz \draw[red, thick, ->] (0,0) -- (0,0.2);~~~\fix{-}} & 3  & \multirow{-2}{*}{\tikz \draw[red, thick, ->] (0,0) -- (0,0.2);~~~\fix{-}} & \fix{1} & \multirow{-2}{*}{\tikz \draw[red, thick, ->] (0,0) -- (0,0.2);~~~-} & \multirow{-2}{*}{\fix{5.13}}\\ \hline
\multicolumn{1}{c|}{}    & \fix{G-A}& 35 &  & 5 &  & 38 &  & 40&  & 66&  &\fix{3} &  & \fix{3} &     & 2  &     & \fix{1} &     &     \\
\multicolumn{1}{c|}{\multirow{-2}{*}{Meta-Llama-3.1}}  & \fix{H} & 41 & \multirow{-2}{*}{\tikz \draw[red, thick, ->] (0,0) -- (0,0.2);~1.17} & 9 & \multirow{-2}{*}{\tikz \draw[red, thick, ->] (0,0) -- (0,0.2);~1.80} & 43 & \multirow{-2}{*}{\tikz \draw[red, thick, ->] (0,0) -- (0,0.2);~1.13} & 81& \multirow{-2}{*}{\tikz \draw[red, thick, ->] (0,0) -- (0,0.2);~2.03} & 150     & \multirow{-2}{*}{\tikz \draw[red, thick, ->] (0,0) -- (0,0.2);~2.27} & \fix{9}  & \multirow{-2}{*}{\tikz \draw[red, thick, ->] (0,0) -- (0,0.2);~\fix{3.00}} & \fix{7} & \multirow{-2}{*}{\tikz \draw[red, thick, ->] (0,0) -- (0,0.2);~\fix{2.33}}    & 5  & \multirow{-2}{*}{\tikz \draw[red, thick, ->] (0,0) -- (0,0.2);~2.50}   & \fix{1} & \multirow{-2}{*}{\fix{1.00}}& \multirow{-2}{*}{\fix{1.91}}\\ \hline
\multicolumn{1}{c|}{}    & \fix{G-A}& 49 &  & 1 &  & 17 &  & 54&  & 100     &  & \fix{3} &  & \fix{1} &     & 4  &     & \fix{0} &     &     \\
\multicolumn{1}{c|}{\multirow{-2}{*}{Mistral-v0.3}}    & \fix{H} & 51 & \multirow{-2}{*}{\tikz \draw[red, thick, ->] (0,0) -- (0,0.2);~1.04} & 1 & \multirow{-2}{*}{~\fix{1.00}}   & 50 & \multirow{-2}{*}{\tikz \draw[red, thick, ->] (0,0) -- (0,0.2);~2.94} & 75& \multirow{-2}{*}{\tikz \draw[red, thick, ->] (0,0) -- (0,0.2);~1.39} & 114     & \multirow{-2}{*}{\tikz \draw[red, thick, ->] (0,0) -- (0,0.2);~1.14} & \fix{11}& \multirow{-2}{*}{\tikz \draw[red, thick, ->] (0,0) -- (0,0.2);~\fix{3.67}} & \fix{5} & \multirow{-2}{*}{\tikz \draw[red, thick, ->] (0,0) -- (0,0.2);~\fix{5.00}}    & 5  & \multirow{-2}{*}{\tikz \draw[red, thick, ->] (0,0) -- (0,0.2);~1.25}    & \fix{0}& \multirow{-2}{*}{~~~\fix{-}}    & \multirow{-2}{*}{\fix{2.05}}\\ \hline
\multicolumn{1}{c|}{}    & \fix{\fix{G-A}}& \fix{45} &  & \fix{4} &  & \fix{20} &  & \fix{53}&  & \fix{68}&  & \fix{7}  &  & \fix{1} &     & \fix{2}  &     & \fix{0} &     &     \\
\multicolumn{1}{c|}{\multirow{-2}{*}{\fix{Meta-Llama-3.3}}}  & \fix{\fix{H}}& \fix{53}& \multirow{-2}{*}{\tikz \draw[red, thick, ->] (0,0) -- (0,0.2);~\fix{1.18}} & \fix{5} & \multirow{-2}{*}{\tikz \draw[red, thick, ->] (0,0) -- (0,0.2);~\fix{1.25}} &\fix{47} & \multirow{-2}{*}{\tikz \draw[red, thick, ->] (0,0) -- (0,0.2);~\fix{2.35}} & \fix{74} & \multirow{-2}{*}{\tikz \draw[red, thick, ->] (0,0) -- (0,0.2);~\fix{1.40}} & \fix{117}     & \multirow{-2}{*}{\tikz \draw[red, thick, ->] (0,0) -- (0,0.2);~\fix{1.72}} & \fix{12} & \multirow{-2}{*}{\tikz \draw[red, thick, ->] (0,0) -- (0,0.2);~\fix{1.71}} & \fix{9} & \multirow{-2}{*}{\tikz \draw[red, thick, ->] (0,0) -- (0,0.2);~\fix{9.00}}    & \fix{10} & \multirow{-2}{*}{\tikz \draw[red, thick, ->] (0,0) -- (0,0.2);~\fix{5.00}}    & \fix{5} & \multirow{-2}{*}{\tikz \draw[red, thick, ->] (0,0) -- (0,0.2);~~~\fix{-}} & \multirow{-2}{*}{\fix{2.95}}\\ \hline
\multicolumn{1}{c|}{}    &\fix{\fix{G-A}}& \fix{53} &  &\fix{5} &  & \fix{11} &  &\fix{61}&  & \fix{46}&  &\fix{6}  &  &\fix{1} &     &\fix{1}  &     &\fix{2} &     &     \\
\multicolumn{1}{c|}{\multirow{-2}{*}{\fix{DeepSeek-V3}}}     &\fix{\fix{H}} &\fix{60} & \multirow{-2}{*}{\tikz \draw[red, thick, ->] (0,0) -- (0,0.2);~\fix{1.13}} & \fix{6} & \multirow{-2}{*}{\tikz \draw[red, thick, ->] (0,0) -- (0,0.2);~\fix{1.20}} & \fix{34}& \multirow{-2}{*}{\tikz \draw[red, thick, ->] (0,0) -- (0,0.2);~\fix{3.09}} & \fix{69}& \multirow{-2}{*}{\tikz \draw[red, thick, ->] (0,0) -- (0,0.2);~\fix{1.13}} & \fix{112} & \multirow{-2}{*}{\tikz \draw[red, thick, ->] (0,0) -- (0,0.2);~\fix{2.43} }&\fix{13}& \multirow{-2}{*}{\tikz \draw[red, thick, ->] (0,0) -- (0,0.2);~\fix{2.17} }& \fix{3}& \multirow{-2}{*}{\tikz \draw[red, thick, ->] (0,0) -- (0,0.2);~\fix{3.00}  }  &\fix{4}  & \multirow{-2}{*}{\tikz \draw[red, thick, ->] (0,0) -- (0,0.2);~\fix{4.00}}    & \fix{4}& \multirow{-2}{*}{\tikz \draw[red, thick, ->] (0,0) -- (0,0.2);~\fix{2.00}}    & \multirow{-2}{*}{\fix{2.24}}\\ \hline
\multicolumn{1}{c|}{}    & \fix{G-A}& 24 && \fix{3}& & 18 & & 39& & 52& & \fix{4}& & \fix{0} & &0  & & \fix{0}& &    \\
\multicolumn{1}{c|}{\multirow{-2}{*}{GPT-4o mini}}     & \fix{H} & 43 & \multirow{-2}{*}{\tikz \draw[red, thick, ->] (0,0) -- (0,0.2);~1.79} & 5 & \multirow{-2}{*}{\tikz \draw[red, thick, ->] (0,0) -- (0,0.2);~1.67} & 59 & \multirow{-2}{*}{\tikz \draw[red, thick, ->] (0,0) -- (0,0.2);~3.28} & 80& \multirow{-2}{*}{\tikz \draw[red, thick, ->] (0,0) -- (0,0.2);~2.05} & 105     & \multirow{-2}{*}{\tikz \draw[red, thick, ->] (0,0) -- (0,0.2);~2.02} &  \fix{15} & \multirow{-2}{*}{\tikz \draw[red, thick, ->] (0,0) -- (0,0.2);~ \fix{3.75}} &  \fix{5} & \multirow{-2}{*}{\tikz \draw[red, thick, ->] (0,0) -- (0,0.2);~~~ \fix{-}} & 3  & \multirow{-2}{*}{\tikz \draw[red, thick, ->] (0,0) -- (0,0.2);~~~-} &  \fix{0} & \multirow{-2}{*}{~~~\fix{-}}    & \multirow{-2}{*}{ \fix{2.22}}\\ \hline
\multicolumn{1}{c|}{}    &  \fix{\fix{G-A}} & \fix{38}  & & \fix{5} &  & \fix{11} & & \fix{45} &  & \fix{64} &  & \fix{2} &  & \fix{1} &  & \fix{0} &  & \fix{2} &  &      \\
\multicolumn{1}{c|}{\multirow{-2}{*}{\fix{GPT-4o}}}   & \fix{\fix{H}} & \fix{40} &  \multirow{-2}{*}{\tikz \draw[red, thick, ->] (0,0) -- (0,0.2);~\fix{1.05}  } & \fix{12} &  \multirow{-2}{*}{\tikz \draw[red, thick, ->] (0,0) -- (0,0.2);~\fix{2.40}} & \fix{37} & \multirow{-2}{*}{\tikz \draw[red, thick, ->] (0,0) -- (0,0.2);~\fix{3.36}} & \fix{54} & \multirow{-2}{*}{\tikz \draw[red, thick, ->] (0,0) -- (0,0.2);~\fix{1.20}} & \fix{96} & \multirow{-2}{*}{\tikz \draw[red, thick, ->] (0,0) -- (0,0.2);~ \fix{1.50}} & \fix{14} & \multirow{-2}{*}{\tikz \draw[red, thick, ->] (0,0) -- (0,0.2);~\fix{7.00}} & \fix{7} &  \multirow{-2}{*}{\tikz \draw[red, thick, ->] (0,0) -- (0,0.2);~\fix{7.00}} & \fix{4} &  \multirow{-2}{*}{\tikz \draw[red, thick, ->] (0,0) -- (0,0.2);~~~\fix{-}} & \fix{4} &  \multirow{-2}{*}{\tikz \draw[red, thick, ->] (0,0) -- (0,0.2);~\fix{2.00}} & \multirow{-2}{*}{\fix{3.19}}\\ \hline
\multicolumn{2}{c|}{Avg. (same Target)}  &  \multicolumn{2}{c|}{\fix{1.56}} &  \multicolumn{2}{c|}{\fix{2.72}} & \multicolumn{2}{c|}{\fix{2.63}} &  \multicolumn{2}{c|}{\fix{1.82}} & \multicolumn{2}{c|}{\fix{2.11}} &\multicolumn{2}{c|}{ \fix{3.76}} & \multicolumn{2}{c|}{ \fix{4.89}} & \multicolumn{2}{c|}{\fix{3.03}} & \multicolumn{2}{c|}{\fix{1.33}} & \fix{2.60 (all)}\\ \bottomrule
\end{tabular}
}
\vspace{-0.05cm}
\begin{tablenotes}    %这行要添加， 从这开始
    \footnotesize   %这行要添加
    \item \fix{G-A is {\baseline} and H is {\toolname}.}
    $P_{uniq}$ is the number of unique hallucinated packages found by {\toolname} or {\baseline}, and $R_{inc}$ indicates the improvement rate of {\toolname} compared to {\baseline}.
\end{tablenotes}
\vspace{-0.7cm}
\label{table:rq2}
\end{table*}
% \usepackage{multirow}
% \usepackage{graphicx}
\begin{table}[]
\centering
\setlength{\tabcolsep}{2pt}
\caption{\fix{RQ1: Multiple Runs Results}}
\vspace{-0.3cm}
% \resizebox{\columnwidth}{!}{\color{blue}%
\resizebox{0.9\columnwidth}{!}{%
\begin{tabular}{c|c|ccc|ccc|ccc}
\hline
\multicolumn{2}{c|}{\multirow{2}{*}{\diagbox[width=11em,trim=l]{Tester}{Target}}} & \multicolumn{3}{c|}{Meta-Llama-3.1} & \multicolumn{3}{c|}{Mistral-v0.3} & \multicolumn{3}{c}{GPT-4o mini} \\ \cline{3-11} 
& & $\mu P_{uniq}$   & $\mu R_{inc}$ & P & $\mu P_{uniq}$   & $\mu R_{inc}$ & P & $\mu P_{uniq}$   & $\mu R_{inc}$ & P \\ \hline
\multirow{2}{*}{Meta-Llama-3} & \fix{G-A}   &39.33 & \multirow{2}{*}{\tikz \draw[red, thick, ->] (0,0) -- (0,0.2);~1.62} & \multirow{2}{*}{0.00} & 54.33 & \multirow{2}{*}{\tikz \draw[red, thick, ->] (0,0) -- (0,0.2);~1.85} & \multirow{2}{*}{0.00} & 2.67 & \multirow{2}{*}{\tikz \draw[red, thick, ->] (0,0) -- (0,0.2);~1.75} & \multirow{2}{*}{0.15} \\
& \fix{H} & 63.67 &  &  & 100.33 &  &  & 4.67 &  &  \\ \hline
\multirow{2}{*}{Qwen2.5-Coder} & \fix{G-A}   & 44.67 & \multirow{2}{*}{\tikz \draw[red, thick, ->] (0,0) -- (0,0.2);~1.57} & \multirow{2}{*}{0.04} & 59.33 & \multirow{2}{*}{\tikz \draw[red, thick, ->] (0,0) -- (0,0.2);~1.69} & \multirow{2}{*}{0.01} & 0.33 & \multirow{2}{*}{\tikz \draw[red, thick, ->] (0,0) -- (0,0.2);~10.00} & \multirow{2}{*}{0.06} \\
& \fix{H} & 70.33 &  &  & 100.00 &  &  & 3.33 &  &  \\ \hline
\multirow{2}{*}{deepseek-coder} & \fix{G-A}   & 27.67 & \multirow{2}{*}{\tikz \draw[red, thick, ->] (0,0) -- (0,0.2);~4.23} & \multirow{2}{*}{0.02} & 36.00 & \multirow{2}{*}{\tikz \draw[red, thick, ->] (0,0) -- (0,0.2);~4.17} & \multirow{2}{*}{0.01} & 0.00 & \multirow{2}{*}{\tikz \draw[red, thick, ->] (0,0) -- (0,0.2);~~~-} & \multirow{2}{*}{0.03} \\
& \fix{H} & 117.00 &  &  & 150.00 &  &  & 4.00 &  &  \\ \hline
\multirow{2}{*}{Meta-Llama-3.1} & \fix{G-A}   & 33.67 & \multirow{2}{*}{\tikz \draw[red, thick, ->] (0,0) -- (0,0.2);~2.28} & \multirow{2}{*}{0.01} & 69.00 & \multirow{2}{*}{\tikz \draw[red, thick, ->] (0,0) -- (0,0.2);~1.91} & \multirow{2}{*}{0.01} & 2.00 & \multirow{2}{*}{\tikz \draw[red, thick, ->] (0,0) -- (0,0.2);~2.67} & \multirow{2}{*}{0.02} \\
& \fix{H} & 76.67 &  &  & 131.67 &  &  & 5.33 &  &  \\ \hline
\multirow{2}{*}{Mistral-v0.3} & \fix{G-A}   & 42.00 & \multirow{2}{*}{\tikz \draw[red, thick, ->] (0,0) -- (0,0.2);~2.06} & \multirow{2}{*}{0.01} & 50.33 & \multirow{2}{*}{\tikz \draw[red, thick, ->] (0,0) -- (0,0.2);~2.43} & \multirow{2}{*}{0.01} & 3.00 & \multirow{2}{*}{\tikz \draw[red, thick, ->] (0,0) -- (0,0.2);~1.44} & \multirow{2}{*}{0.16} \\
& \fix{H} & 86.67 &  &  & 122.33 &  &  & 4.33 &  &  \\ \hline
\multirow{2}{*}{Meta-Llama-3.3} & \fix{G-A}   & 53.67 & \multirow{2}{*}{\tikz \draw[red, thick, ->] (0,0) -- (0,0.2);~1.30} & \multirow{2}{*}{0.01} & 74.67 & \multirow{2}{*}{\tikz \draw[red, thick, ->] (0,0) -- (0,0.2);~1.62} & \multirow{2}{*}{0.00} & 2.33 & \multirow{2}{*}{\tikz \draw[red, thick, ->] (0,0) -- (0,0.2);~3.71} & \multirow{2}{*}{0.01} \\
& \fix{H} & 70.00 &  &  & 121.00 &  &  & 8.67 &  &  \\ \hline
\multirow{2}{*}{DeepSeek-V3} & \fix{G-A}   & 43.67 & \multirow{2}{*}{\tikz \draw[red, thick, ->] (0,0) -- (0,0.2);~1.60} & \multirow{2}{*}{0.05} & 60.00 & \multirow{2}{*}{\tikz \draw[red, thick, ->] (0,0) -- (0,0.2);~1.76} & \multirow{2}{*}{0.01} & 0.33 & \multirow{2}{*}{\tikz \draw[red, thick, ->] (0,0) -- (0,0.2);~14.00} & \multirow{2}{*}{0.00} \\
& \fix{H} & 70.00 &  &  & 105.33 &  &  & 4.67 &  &  \\ \hline
\multirow{2}{*}{GPT-4o mini} & \fix{G-A}   & 40.67 & \multirow{2}{*}{\tikz \draw[red, thick, ->] (0,0) -- (0,0.2);~1.77} & \multirow{2}{*}{0.00} & 58.33 & \multirow{2}{*}{\tikz \draw[red, thick, ->] (0,0) -- (0,0.2);~1.86} & \multirow{2}{*}{0.00} & 0.67 & \multirow{2}{*}{\tikz \draw[red, thick, ->] (0,0) -- (0,0.2);~5.00} & \multirow{2}{*}{0.04} \\
& \fix{H} & 72.00 &  &  & 108.67 &  &  & 3.33 &  &  \\ \hline
\multirow{2}{*}{GPT-4o} & \fix{G-A}   & 48.33 & \multirow{2}{*}{\tikz \draw[red, thick, ->] (0,0) -- (0,0.2);~1.14} & \multirow{2}{*}{0.03} & 54.33 & \multirow{2}{*}{\tikz \draw[red, thick, ->] (0,0) -- (0,0.2);~2.09} & \multirow{2}{*}{0.01} & 0.00 & \multirow{2}{*}{\tikz \draw[red, thick, ->] (0,0) -- (0,0.2);~~~-} & \multirow{2}{*}{0.09} \\
& \fix{H} & 55.00 &  &  & 113.67 &  &  & 4.67 &  &  \\ \hline
\multicolumn{2}{c|}{Avg.(Same Target)} & \multicolumn{3}{c|}{1.95}&\multicolumn{3}{c|}{2.15}&\multicolumn{3}{c}{5.51}\\ \hline
\multicolumn{2}{c|}{$\mu CV$ (H/G-A)} & \multicolumn{3}{c|}{0.10/0.17}&\multicolumn{3}{c|}{0.11/0.14}&\multicolumn{3}{c}{0.39/0.72}\\ \hline
\end{tabular}%
}
\begin{tablenotes}    %这行要添加， 从这开始
    \footnotesize   %这行要添加
    \item \fix{$\mu P_{uniq}$ and $\mu R_{inc}$ are the means of multiple runs results. $P$ is the value of Welch's t-test. $\mu CV$ is the mean of the coefficient of variation.}
\end{tablenotes}
\vspace{-0.6cm}
\label{table:rq21}
\end{table}
\noindent \textbf{Metric.}
% In RQ1, we analyze the ``Package Hallucination Rate'' (PHR), which is defined as the proportion of model responses that contain at least one hallucinated package~\cite{krishna2025importing}, to evaluate whether coding tasks generated by {\toolname} can effectively trigger LLM's package hallucinations.
% We calculate PHR by checking the number of rounds in which hallucinated packages appeared.
% Additionally, similar to unique bugs, which is a widely used metric in fuzzing, we use the number of unique hallucinated packages to evaluate {\toolname}'s capability of testing LLMs for package hallucinations (RQ2 and RQ3).
\fix{We evaluate the effectiveness of {\toolname} using the Package Hallucination Rate (PHR) and the number of unique hallucinated packages (RQ1 and RQ3).
PHR is the proportion of model responses containing hallucinated packages over the total number of responses~\cite{krishna2025importing}, and is used to assess whether the coding tasks generated by {\toolname} can trigger package hallucinations.
The number of unique hallucinated packages is analogous to a widely used metric in fuzzing (i.e., unique bugs) and is used to evaluate {\toolname}'s ability to test LLMs for package hallucinations.
We use the Diversity Index to assess task diversity (RQ2 and RQ3).
The Diversity Index is defined as the total of clusters and noise points.}
% perform clustering analysis on the generated tasks and use the sum of the number of clusters and noise points as a metric to quantitatively assess diversity.}

% \input{table/RQ2}
\noindent \textbf{Environment.}
Our experiments are conducted on a server with four NVIDIA A800 GPUs.
The server has an Intel Xeon Platinum 8358P CPU with 32 cores and 2TB of memory.
The version of vLLM~\cite{vllm} used is v0.6.1.

\subsection{RQ1: The Capability in Testing LLMs}

\label{subsec:rq1}
\fix{To answer RQ1, we calculate the number of unique hallucinated packages found by {\toolname} and {\baseline}, as well as the PHR for different target models.
We select the top 100 Python packages from libraries.io~\cite{libraries} according to the ``SourceRank'' and collect their information.
Each run is limited to 1,000 rounds.
To ensure comprehensiveness, we use nine models as tester and target models, forming 81 model combinations.}
The temperature of the tester models is fixed at 0 for determinism, whereas the target model’s temperature is set to 0.7 to balance creativity and stability~\cite{chen2023chatgpt}. 
The maximum token limit is set to 3,000.

\begin{figure}[t]
\centerline{\includegraphics[width=0.95\linewidth]{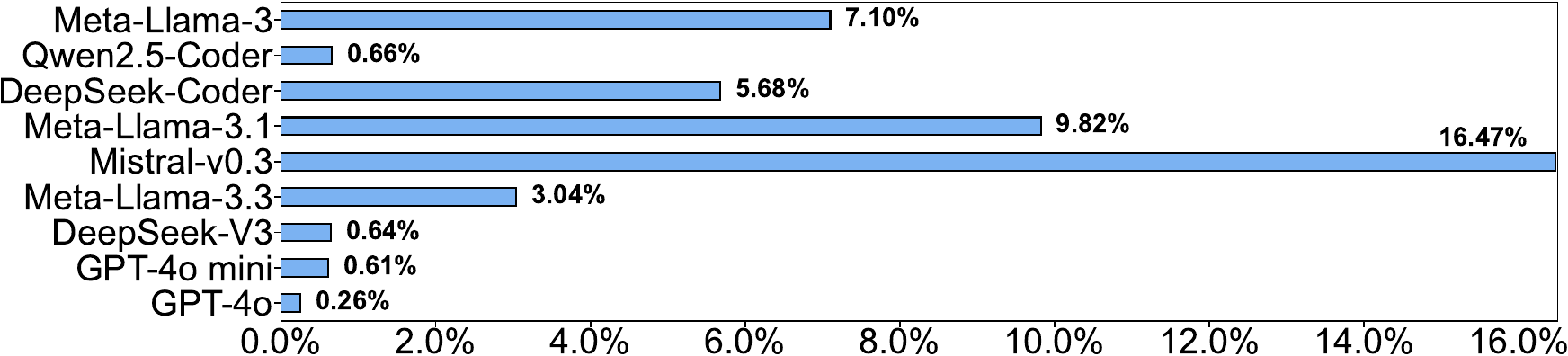}}
\vspace{-0.3cm}
\caption{RQ1: Average PHR of Different Target Models}
\label{average}
\vspace{-0.5cm}
\end{figure}
\begin{figure}[t]
\centerline{\includegraphics[width=0.95\linewidth]{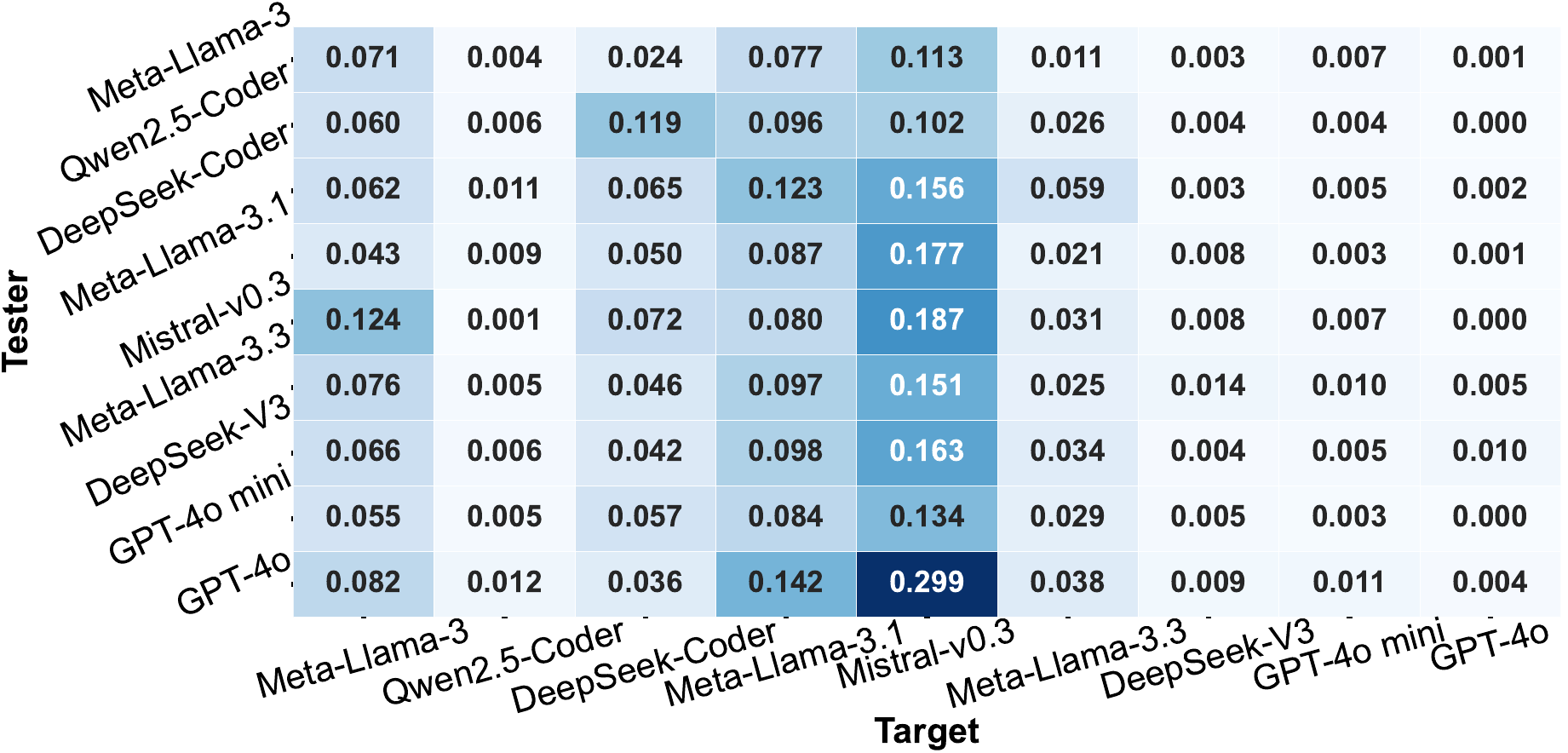}}
\vspace{-0.3cm}
\caption{RQ1: Heatmap of PHR with Different Models}
\label{hmap}
\vspace{-0.7cm}
\end{figure}

\fix{
Table~\ref{table:rq2} reports the results of unique hallucinated packages.  
Overall, {\toolname} outperforms {\baseline} across most model combinations, with an average number of unique hallucinated packages increasing by 2.60×.}
% \fix{
% Figure~\ref{average} shows the average PHR for each target model, which is calculated by averaging the PHR under different tester models.
% GPT-4o, GPT-4o mini, DeepSeek-V3, and Qwen2.5-Coder show the lower PHR, only 0.26\%, 0.50\%, 0.64\% and 0.66\%, while Mistral-v0.3 shows the highest PHR, 16.47\%.
% PHR of DeepSeek-Coder, Meta-Llama-3, and Meta-Llama-3.1 are 5.68\%, 7.10\%, and 9.82\%.
% Figure~\ref{hmap} further shows the PHR of target models under different model combinations.
% The results indicate that, in most model combinations, {\toolname} can trigger package hallucinations in the target models.
% The above results demonstrate the effectiveness of the tasks generated by {\toolname} in triggering package hallucinations.
% To analyze the capability of {\toolname} in testing LLMs for package hallucinations, we count the number of unique hallucinated packages for {\toolname} and {\baseline} across different model combinations, which is shown in Table~\ref{table:rq2}.
% For most combinations of tester and target models, {\toolname} outperforms {\baseline}, identifying more unique hallucinated packages, with the average number of unique hallucinated packages increasing by 2.57x.}
\fix{We combine Table~\ref{table:rq2} and Figure~\ref{average} to analyze the results on different target models.
Figure~\ref{average} shows the average PHR for each target model, computed by averaging the PHR under different tester models.
{\toolname} triggers package hallucinations across all target models.
GPT-4o, GPT-4o mini, DeepSeek-V3, and Qwen2.5-Coder exhibit relatively low PHRs (0.26–0.66\%), while Mistral-v0.3 shows the highest (16.47\%).  
The PHR for DeepSeek-Coder, Meta-Llama-3, and Meta-Llama-3.1 are 5.68\%, 7.10\%, and 9.82\%.  
Compared with {\baseline}, {\toolname} finds more unique hallucinated packages across all target models, demonstrating its applicability.
The largest improvement is achieved when the target model is DeepSeek-V3 (4.89x).
Additionally, when GPT-4o mini and GPT-4o are used as target models, {\toolname} outperforms {\baseline}, highlighting its practical value.}
\fix{To analyze the impact of different tester models, we analyze the results using different tester models, as shown in Table~\ref{table:rq2} and Figure~\ref{hmap}. 
For the number of unique hallucinated packages, {\toolname} outperforms {\baseline} under all tester models, with the largest improvement (5.13×) achieved when the tester is DeepSeek-Coder.
Moreover, {\toolname} triggers hallucinations in most model combinations, and no tester model consistently outperforms the others.
This suggests that the performance of {\toolname} does not strongly depend on the specific tester model, and it remains effective even with smaller-scale LLMs.}

\fix{To evaluate stability,  we follow MetaQA~\cite{yang2025hallucination}, select GPT-4o mini, Meta-Llama-3.1, and Mistral-v0.3 as target models.
For each target, we use all tester models and run {\toolname} and {\baseline} three times.
Following prior studies~\cite{liyanage2025assessing, shriver2019evaluating}, we use Welch's t-test to assess statistical significance and report the coefficients of variation (CV)~\cite{abdi2010coefficient}.
As shown in Table~\ref{table:rq21}, {\toolname} yields an average 3.02× improvement over {\baseline}.
This improvement is statistically significant (p$<$0.05) for most model combinations, and {\toolname} exhibits consistently lower CV, indicating greater stability.
For some combinations, improvements are not significant due to the low hallucination rates of the target models.
The low hallucination rates result in limited differences in means relative to variances, leading to smaller effect sizes in the Welch's t-test and larger CV compared to other target models.}

\begin{table*}[t]
\vspace{-0.3cm}
\centering
\setlength{\abovecaptionskip}{0.05cm} %# 调整间距
\setlength{\belowcaptionskip}{0.1cm}
\caption{RQ3: Unique Hallucinated Packages Results of Ablation}
\vspace{-0.1cm}
\resizebox{0.75\textwidth}{!}{
\begin{tabular}{cc|c|cl|cl|cl|cl}
\hline
\multicolumn{1}{c|}{\multirow{2}{*}{\begin{tabular}[c]{@{}c@{}}Tester Model\end{tabular}}} & \multirow{2}{*}{\begin{tabular}[c]{@{}c@{}}Target Model\end{tabular}} & {\toolname} & \multicolumn{2}{c|}{ W/O Expansion} & \multicolumn{2}{c|}{ W/O Power} & \multicolumn{2}{c|}{ W/O Phrase} & \multicolumn{2}{c}{ W/O All} \\ \cline{3-11} 
\multicolumn{1}{c|}{}                                                                          &                                                                         & $P_{uniq}$                  & $P_{uniq}$                     & $R_{inc}$                    & $P_{uniq}$                   & $R_{inc}$                  & $P_{uniq}$                         & $R_{inc}$             & $P_{uniq}$                 & $R_{inc}$                 \\ \hline
\multicolumn{1}{c|}{\multirow{2}{*}{Meta-Llama-3}}                                             & Meta-Llama-3.1                                                          & 65                           & 48                              & {\tikz \draw[red, thick, ->] (0,0) -- (0,0.2);~1.35}                         & 44                            & {\tikz \draw[red, thick, ->] (0,0) -- (0,0.2);~1.48}                     & 43                                  & {\tikz \draw[red, thick, ->] (0,0) -- (0,0.2);~1.51}                   & 1                           & {\tikz \draw[red, thick, ->] (0,0) -- (0,0.2);~65.00 }                     \\
\multicolumn{1}{c|}{}                                                                          & Mistral-v0.3                                                            & 95                           & 80                              & {\tikz \draw[red, thick, ->] (0,0) -- (0,0.2);~1.19    }                      & 31                            & {\tikz \draw[red, thick, ->] (0,0) -- (0,0.2);~3.06      }                  & 58                                  & {\tikz \draw[red, thick, ->] (0,0) -- (0,0.2);~1.64      }             & 1                           & {\tikz \draw[red, thick, ->] (0,0) -- (0,0.2);~95.00  }                    \\ \hline
\multicolumn{1}{c|}{\multirow{2}{*}{Qwen2.5-Coder}}                                            & Meta-Llama-3.1                                                          & 72                           & 74                              & {\tikz \draw[red, thick, ->] (0,0) -- (0,0.2);~1.26   }                       & 56                            & {\tikz \draw[red, thick, ->] (0,0) -- (0,0.2);~1.29    }                    & 32                                  & {\tikz \draw[red, thick, ->] (0,0) -- (0,0.2);~2.25        }           & 0                           & {\tikz \draw[red, thick, ->] (0,0) -- (0,0.2);~~~- }                         \\
\multicolumn{1}{c|}{}                                                                          & Mistral-v0.3                                                            & 86                           & 86                              & {~\fix{1.00}}                      & 59                            & {\tikz \draw[red, thick, ->] (0,0) -- (0,0.2);~1.46 }                       & 48          & {\tikz \draw[red, thick, ->] (0,0) -- (0,0.2);~1.79         }          & 6                           & {\tikz \draw[red, thick, ->] (0,0) -- (0,0.2);~14.33}                       \\ \hline
\multicolumn{1}{c|}{\multirow{2}{*}{DeepSeek-Coder}}                                           & Meta-Llama-3.1                                                          & 114                          & 74                              & {\tikz \draw[red, thick, ->] (0,0) -- (0,0.2);~1.54 }                         & 76                            & {\tikz \draw[red, thick, ->] (0,0) -- (0,0.2);~1.50}                        & 31                                  & {\tikz \draw[red, thick, ->] (0,0) -- (0,0.2);~3.68     }              & 1                           & {\tikz \draw[red, thick, ->] (0,0) -- (0,0.2);~114.00}                      \\
\multicolumn{1}{c|}{}                                                                          & Mistral-v0.3                                                            & 184                          & 87                              & {\tikz \draw[red, thick, ->] (0,0) -- (0,0.2);~2.11 }                         & 143                           & {\tikz \draw[red, thick, ->] (0,0) -- (0,0.2);~1.29 }                       & 56                                  & {\tikz \draw[red, thick, ->] (0,0) -- (0,0.2);~3.29  }                 & 1                           & {\tikz \draw[red, thick, ->] (0,0) -- (0,0.2);~184.00       }               \\ \hline
\multicolumn{1}{c|}{\multirow{2}{*}{Meta-Llama-3.1}}                                           & Meta-Llama-3.1                                                          & 81                           & 66                              & {\tikz \draw[red, thick, ->] (0,0) -- (0,0.2);~1.23    }                      & 59                           & {\tikz \draw[red, thick, ->] (0,0) -- (0,0.2);~1.37}                        & 44                                  & {\tikz \draw[red, thick, ->] (0,0) -- (0,0.2);~1.84    }               & 8                           & {\tikz \draw[red, thick, ->] (0,0) -- (0,0.2);~10.13   }                    \\
\multicolumn{1}{c|}{}                                                                          & Mistral-v0.3                                                            & 150                          & 119                             & {\tikz \draw[red, thick, ->] (0,0) -- (0,0.2);~1.26}                          & 134                           & {\tikz \draw[red, thick, ->] (0,0) -- (0,0.2);~1.12 }                       & 85                                  & {\tikz \draw[red, thick, ->] (0,0) -- (0,0.2);~1.76 }                  & 0                           & {\tikz \draw[red, thick, ->] (0,0) -- (0,0.2);~~~- }                         \\ \hline
\multicolumn{1}{c|}{\multirow{2}{*}{Mistral-v0.3}}                                             & Meta-Llama-3.1                                                          & 75                           & 73                              & {\tikz \draw[red, thick, ->] (0,0) -- (0,0.2);~1.03    }                      & 69                            & {\tikz \draw[red, thick, ->] (0,0) -- (0,0.2);~1.09 }                       & 42                                  & {\tikz \draw[red, thick, ->] (0,0) -- (0,0.2);~1.79 }                  & 0                           & {\tikz \draw[red, thick, ->] (0,0) -- (0,0.2);~~~-   }                       \\
\multicolumn{1}{c|}{}                                                                          & Mistral-v0.3                                                            & 114                          & 88                              & {\tikz \draw[red, thick, ->] (0,0) -- (0,0.2);~1.30 }                            & 98                            &{\tikz \draw[red, thick, ->] (0,0) -- (0,0.2);~1.16}                           & 60                                  & {\tikz \draw[red, thick, ->] (0,0) -- (0,0.2);~1.90}                      & 2                           & {\tikz \draw[red, thick, ->] (0,0) -- (0,0.2);~57.00 }                        \\ \hline
\multicolumn{1}{c|}{\multirow{2}{*}{\fix{Meta-Llama-3.3}}}                                              & \fix{Meta-Llama-3.1}                                                          & \fix{74}                          & \fix{63}                              & {\tikz \draw[red, thick, ->] (0,0) -- (0,0.2);~\fix{1.17} }                            & \fix{45}                            & {\tikz \draw[red, thick, ->] (0,0) -- (0,0.2);~\fix{1.64}       }                   & \fix{36}                                  & {\tikz \draw[red, thick, ->] (0,0) -- (0,0.2);~\fix{2.06}     }                 & \fix{1}                           & {\tikz \draw[red, thick, ->] (0,0) -- (0,0.2);~\fix{74.00}   }                      \\
\multicolumn{1}{c|}{}                                                                          & \fix{Mistral-v0.3}                                                            & \fix{117}                          & \fix{113}                              & {\tikz \draw[red, thick, ->] (0,0) -- (0,0.2);~\fix{1.04} }                            & \fix{77}                           & {\tikz \draw[red, thick, ->] (0,0) -- (0,0.2);~\fix{1.52}        }                   & \fix{39}                                  & {\tikz \draw[red, thick, ->] (0,0) -- (0,0.2);~\fix{3.00}    }                  & \fix{1}                          & {\tikz \draw[red, thick, ->] (0,0) -- (0,0.2);~\fix{117.00}    }                      \\ \hline
\multicolumn{1}{c|}{\multirow{2}{*}{\fix{DeepSeek-V3}}}                                              & \fix{Meta-Llama-3.1}                                                          & \fix{69}                          & \fix{57}                              & {\tikz \draw[red, thick, ->] (0,0) -- (0,0.2);~\fix{1.21} }                            & \fix{60}                            & {\tikz \draw[red, thick, ->] (0,0) -- (0,0.2);~\fix{1.15}        }                   & \fix{59}                                 & {\tikz \draw[red, thick, ->] (0,0) -- (0,0.2);~\fix{1.17}     }                 & \fix{14}                           & {\tikz \draw[red, thick, ->] (0,0) -- (0,0.2);~\fix{4.93}  }                      \\
\multicolumn{1}{c|}{}                                                                          & \fix{Mistral-v0.3}                                                            & \fix{112}                          & \fix{94}                              & {\tikz \draw[red, thick, ->] (0,0) -- (0,0.2);~\fix{1.19} }                            & \fix{97}                           & {\tikz \draw[red, thick, ->] (0,0) -- (0,0.2);~\fix{1.15}        }                   & \fix{52}                                  & {\tikz \draw[red, thick, ->] (0,0) -- (0,0.2);~\fix{2.15}    }                  & \fix{11}                          & {\tikz \draw[red, thick, ->] (0,0) -- (0,0.2);~\fix{10.18}    }                      \\ \hline
\multicolumn{1}{c|}{\multirow{2}{*}{GPT-4o mini}}                                              & Meta-Llama-3.1                                                          & 80                          & 56                              & {\tikz \draw[red, thick, ->] (0,0) -- (0,0.2);~1.43 }                            & 52                            & {\tikz \draw[red, thick, ->] (0,0) -- (0,0.2);~1.54        }                   & 36                                  & {\tikz \draw[red, thick, ->] (0,0) -- (0,0.2);~2.22     }                 & 1                           & {\tikz \draw[red, thick, ->] (0,0) -- (0,0.2);~80.00   }                      \\
\multicolumn{1}{c|}{}                                                                          & Mistral-v0.3                                                            & 105                          & 86                              & {\tikz \draw[red, thick, ->] (0,0) -- (0,0.2);~1.22 }                            & 100                           & {\tikz \draw[red, thick, ->] (0,0) -- (0,0.2);~1.05        }                   & 53                                  & {\tikz \draw[red, thick, ->] (0,0) -- (0,0.2);~1.98    }                  & 15                          & {\tikz \draw[red, thick, ->] (0,0) -- (0,0.2);~7.00    }                      \\ \hline
\multicolumn{1}{c|}{\multirow{2}{*}{{\fix{GPT-4o}}}}                                              & \fix{Meta-Llama-3.1}                                                          & \fix{54}                          & \fix{53}                              & {\tikz \draw[red, thick, ->] (0,0) -- (0,0.2);~\fix{1.02} }                            & \fix{53}                           & {\tikz \draw[red, thick, ->] (0,0) -- (0,0.2);~\fix{1.02}       }                   & \fix{42}                                  & {\tikz \draw[red, thick, ->] (0,0) -- (0,0.2);~\fix{1.29}     }                 & \fix{5}                           & {\tikz \draw[red, thick, ->] (0,0) -- (0,0.2);~\fix{10.80}   }                      \\
\multicolumn{1}{c|}{}                                                                          & \fix{Mistral-v0.3}                                                            & \fix{96}                          & \fix{77}                              & {\tikz \draw[red, thick, ->] (0,0) -- (0,0.2);~\fix{1.25} }                            & \fix{91}                           & {\tikz \draw[red, thick, ->] (0,0) -- (0,0.2);~\fix{1.05}        }                   & \fix{57}                                 & {\tikz \draw[red, thick, ->] (0,0) -- (0,0.2);~\fix{1.68}    }                  & \fix{14}                          & {\tikz \draw[red, thick, ->] (0,0) -- (0,0.2);~\fix{6.86}    }                      \\ \hline
\multicolumn{2}{c|}{\fix{Avg. $R_{DI}$}}                                                                                                                            &   -                 & \multicolumn{2}{c|}{\fix{1.26}}                                       & \multicolumn{2}{c|}{\fix{1.25}}                                   & \multicolumn{2}{c|}{\fix{4.40}}                                    & \multicolumn{2}{c}{\fix{116.46}}                              \\ \hline
\multicolumn{2}{c|}{Avg. $R_{inc}$}                                                                                                                             &   -                 & \multicolumn{2}{c|}{\fix{1.27}}                                       & \multicolumn{2}{c|}{\fix{1.39}}                                   & \multicolumn{2}{c|}{\fix{2.06}}                                    & \multicolumn{2}{c}{\fix{56.68}  }                              \\ \hline
\end{tabular}
}
\begin{tablenotes}    %这行要添加， 从这开始
    \footnotesize               %这行要添加
    \item \fix{Avg. $R_{DI}$ is the mean improvement ({\toolname}/variant) across all model combinations and DBSCAN parameter settings.}
\end{tablenotes}
\vspace{-0.5cm}
\label{table:rq3}
\end{table*}

Further analysis shows a common phenomenon across most models: the tendency to recommend packages composed of technical terms mentioned in the task.
This phenomenon is particularly pronounced when the hallucinated package exists in another programming language.
For instance, DeepSeek-Coder tends to generate the hallucinated package \code{jsonwebtoken} when responding to tasks involving ``JSON Web Tokens'', and both Meta-Llama-3.1 and Mistral-v0.3 tend to generate the hallucinated package \code{apache-arrow} when responding to tasks involving ``Apache Arrow''.
\fix{We also investigate the distribution of hallucinated packages across different models.
Our analysis reveals that there are 190 non-existent packages and 176 other-language packages recommended by multiple models.
Among them, the package} \code{pkg_resources} \fix{is recommended by all models, and 17 packages are recommended by more than half of the models.
This indicates that the same hallucinated packages exist across different models.
Considering the architectural differences between the models, we believe that most of these hallucinated packages likely originate from the training data.}

Qwen2.5-Coder shows a low PHR in our evaluation, which is inconsistent with the results of other studies~\cite{krishna2025importing,spracklen2024we}.
\fix{We find that Qwen2.5-Coder often refuses to generate code even when the same task can be responded to by other models, which limits the effectiveness of both {\toolname} and {\baseline}.
To further investigate this issue, we query other models of Qwen using the same tasks that triggered refusals in Qwen2.5-Coder.
Several models in the Qwen2.5-code series also refuse to respond, whereas the Qwen2.5 series and newer Qwen3/Qwen3-code models respond normally.
This behavior is similar to the over-refusal phenomenon reported in the previous study~\cite{cui2024or}.
Since it appears confined to specific models, we consider our results sufficient to validate {\toolname}’s effectiveness.
Additionally, as over-refusal is beyond the scope of this paper, we leave further investigation for future work.}
% Since these results only appeared in specific models and occurred in both {\toolname} and {\baseline}, the current findings are already sufficient to demonstrate the effectiveness of {\toolname}.
% Given that it is difficult to identify the root cause without access to the training data and internal mechanisms of the models, we plan to explore the underlying reasons in the future.
\fix{We also find that there are more unique hallucinated packages in Llama 3.1 compared to Llama 3.
Further analysis of these hallucinated packages reveals that those beginning with }\code{google-cloud} \fix{are widely recommended, especially when the code contains a statement }\code{from google.protobuf import xx}\fix{.
We suspect that this issue is related to the training data.
Additionally, even after removing these specific packages, Llama 3.1 still recommends more hallucinated packages compared to Llama 3.
These results show an unexpected observation from specific cases: although Llama 3.1 generally outperforms Llama 3 in tasks such as code generation, this does not necessarily correspond to a lower hallucination rate.
A similar observation is also found by Spracklen et al.~\cite{spracklen2024we} (CodeLlama 34B and CodeLlama 13B).
Note that whether this observation holds for more models requires broader experiments, which we plan to investigate further in the future.}

% todo 位置调整
\greyboxb{\fix{Summary for RQ1:}} {
\fix{The tasks generated by {\toolname} trigger package hallucinations on multiple models. Compared with {\baseline}, {\toolname} finds more unique hallucinated packages, with the average number increasing by 2.60×. Further analysis reveals that: (1) most models recommend hallucinated packages consisting of technical terms; (2) some hallucinated packages appear across different models.}
}

\subsection{\fix{RQ2: The Diversity of Generated Tasks}}
\label{subsec:rq2}

\fix{To answer RQ2, we employ the model text-embedding-3-small~\cite{embedding} to transform the tasks generated in Section~\ref{subsec:rq1} into high-dimensional vectors, followed by clustering with the DBSCAN algorithm~\cite{schubert2017dbscan}.
We set the neighborhood radius parameter ($\varepsilon$) to ${0.1,0.2,0.3}$ and the minimum samples parameter (\textit{minS}) to ${1,3,5}$ to explore the impact of different parameter settings.
For each setting, we compute the Diversity Index for every model combination and report the average Diversity Index across all combinations.}

\begin{figure}[t]
\centerline{\includegraphics[width=0.99\linewidth]{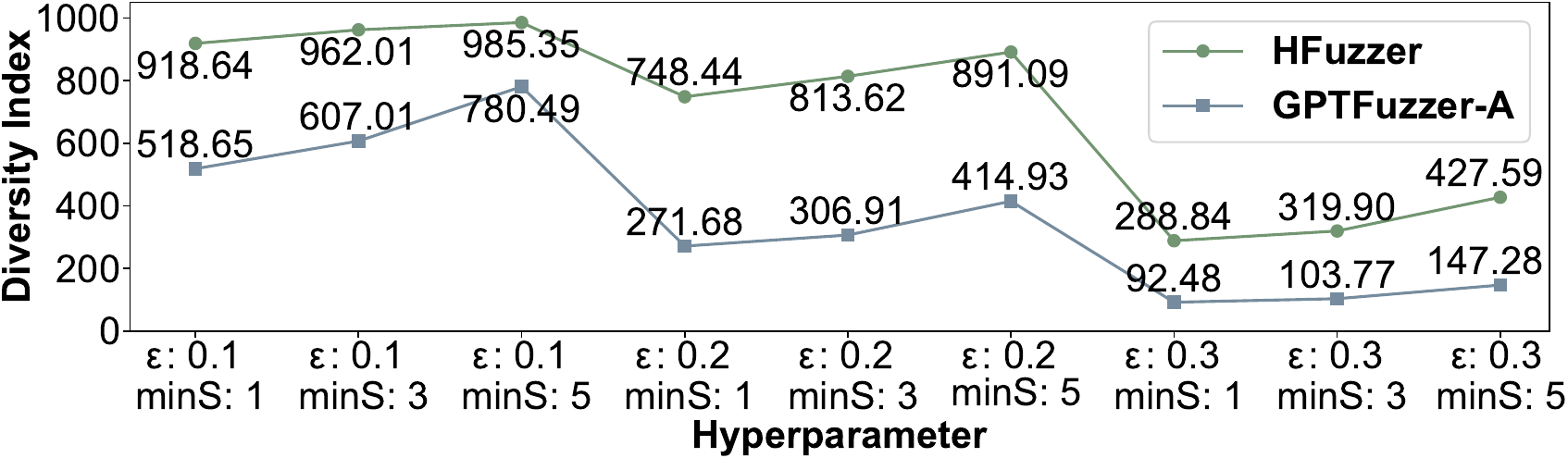}}
\vspace{-0.3cm}
\caption{\fix{RQ2: Average Diversity index of tasks generated under different DBSCAN parameter settings ($\varepsilon \in {0.1,0.2,0.3}$, \textit{minS} $\in {1,3,5}$).}}
\label{rq2_figure}
\vspace{-0.6cm}
\end{figure}

\fix{Across all parameter settings, the tasks generated by {\toolname} consistently have higher Diversity Indices than those generated by {\baseline}, with an average improvement of 2.36×, as shown in Figure~\ref{rq2_figure}.
The Diversity Index decreases as $\varepsilon$ increases, because a larger $\varepsilon$ merges more tasks into the same cluster.
In contrast, increasing \textit{minS} leads to higher Diversity Index, as originally small clusters are split and more tasks are treated as noise points.
Further analysis shows that across all model combinations, tasks generated by {\toolname} exhibit greater diversity compared to those generated by {\baseline}.
For different tester models, the average CV of the Diversity Index is 0.15 for {\toolname} and 0.29 for {\baseline}. 
When $\varepsilon=0.1$, $\varepsilon=0.2$, and $\varepsilon=0.3$, the corresponding CVs are (0.03/0.16), (0.11/0.24), and (0.32/0.47), respectively.
For different target models, the CVs are always around 0.1.
For multiple runs in RQ1, the tasks generated by {\toolname} exhibit a significant improvement in diversity across all parameter settings, with an average Diversity Index increase of 2.46x ($p < 0.05$ and $CV \approx 0.01$).}

\greyboxb{\fix{Summary for RQ2:}} {
\fix{{\toolname} consistently yields more diverse coding tasks than {\baseline}.}}
\subsection{RQ3: The Impact of Different Modules}

\fix{To evaluate the impact of different modules, we design four variants with different modules removed:
\begin{itemize}
    \item \textit{w/o Expansion}: removes the Seed Pool Expansion module introduced in Section~\ref{sec:extract};  
    \item \textit{w/o Power}: removes the Power module and uses random selection;  
    \item \textit{w/o Phrase}: removes the Phrase module and directly uses package descriptions;  
    \item \textit{w/o All}: removes all modules and relies only on the LLM to generate coding tasks.  
\end{itemize}
We compare {\toolname} with these variants to analyze how each model improves {\toolname}.
We use two target models (Mate-Llama-3.1 and Mistral-v0.3) that recommended the largest number of unique hallucinated packages in RQ1 to clearly reveal the impact, and use all tester models. 
Other settings are consistent with RQ1.}

The results are shown in Table~\ref{table:rq3}.
\fix{Each module enhances the performance of {\toolname}.
The Phrase module (\textit{w/o Phrase}) has the largest impact, improving the number of unique hallucinated packages by 2.06× and the diversity Index by 4.40×.}
\fix{The Power module (\textit{w/o Power}) and the Seed Pool Expansion module (\textit{w/o Expansion}) improve the number of unique hallucinated packages by 1.39× and 1.27×, respectively, and the diversity Index by 1.25× and 1.26×.
\textit{w/o All} performs poorly, as relying solely on the LLM often leads to repetitive coding tasks, highlighting the importance of guiding the LLM.}

\greyboxb{Summary for RQ3:} {
\fix{For the number of unique hallucinated packages, each module brings improvement of 1.27x, 1.39x, and 2.06x. For the Diversity Index, each module brings improvements of 1.26x, 1.25x, and 4.40x.}}

\section{Case Study}

We use {\toolname} to test the model GPT-4o~\cite{chatgpt}, which is widely applied across various fields, observe its results, and conduct an in-depth analysis to inspire subsequent studies.
% GPT-4o is widely applied across various fields, with many developers leveraging it to assist in development.
For large-scale evaluation, we use the information of the top 1,000 packages as input and run 10,000 rounds.
We use GPT-4o mini as the tester model to reduce costs.

\begin{figure}[t]
\centerline{\includegraphics[width=0.99\linewidth]{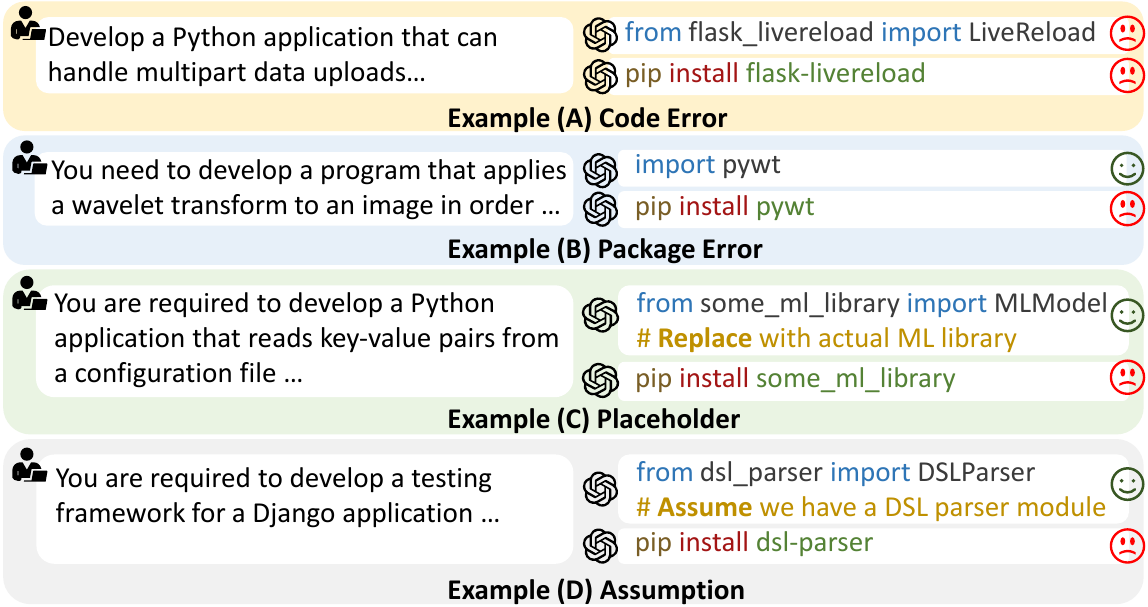}}
\vspace{-0.3cm}
\caption{Examples of Case Study}
\label{ExamplesA}
\vspace{-0.4cm}
\end{figure}

We manually inspect hallucinated packages and conduct a detailed analysis. 
{\toolname} finds 46 unique hallucinated packages, 11 of which are other-language packages.
\fix{Examining the intermediate results reveals two types of hallucinated packages: \textbf{code error} and \textbf{package} \textbf{error}.}
\textbf{Code error} occurs when the generated code contains incorrect import statements.
As shown in Figure~\ref{ExamplesA}~(A), the model is prompted to develop a Python application for handling multipart data uploads from a web interface.
The generated code includes \code{from flask_livereload import LiveReload}, attempting to import a non-existent package. 
By analyzing the code, we find that the intended package is \code{livereload}, but the model produces an incorrect import statement.
\textbf{Package error} occurs when the import statement is correct, but the model returns hallucinated packages in the installation command.
Figure~\ref{ExamplesA}~(B) illustrates this with an example in which the model is required to apply wavelet transforms to images for contour computation.
In the generated code, the model generates the correct import statement \code{import pywt} to import the package \code{PyWavelets}.
However, when providing the installation command, the model incorrectly suggests using \code{pip install pywt}, which indicates that the model cannot correctly match the installation command with the import statement.
\fix{The preliminary examination is performed by one author.
To classify all hallucinated packages, two authors independently classify hallucinated packages as either code errors or package errors.
The classification is consistent due to the clear criteria: after installing the correct package, if the corresponding code can parse this package, it is classified as a package error; otherwise, it is a code error.}
Our classification results show that 34 hallucinated packages belong to package error, while only 12 belong to code error.
\fix{This suggests a potential hypothesis: current LLMs are more prone to package hallucinations when assisting with environment configuration.}
Existing studies~\cite{latendresse2024chatgpt,krishna2025importing} use regular expressions to extract packages, overlooking potential hallucinations that can arise from this process.
Due to the limitations of regular expressions (e.g., the inability to handle Python aliases), automated environment configuration often relies on LLM implementation in real-world scenarios~\cite{hu2025llm}.
% Further research is required to mitigate package hallucinations during the environment configuration phase, enhance LLMs' performance, and defend against potential attacks.
\fix{Whether this hypothesis holds across different models requires further investigation, but it is noteworthy as it highlights a potential scenario influenced by package hallucinations.}

\begin{figure}[t]
\centerline{\includegraphics[width=0.99\linewidth]{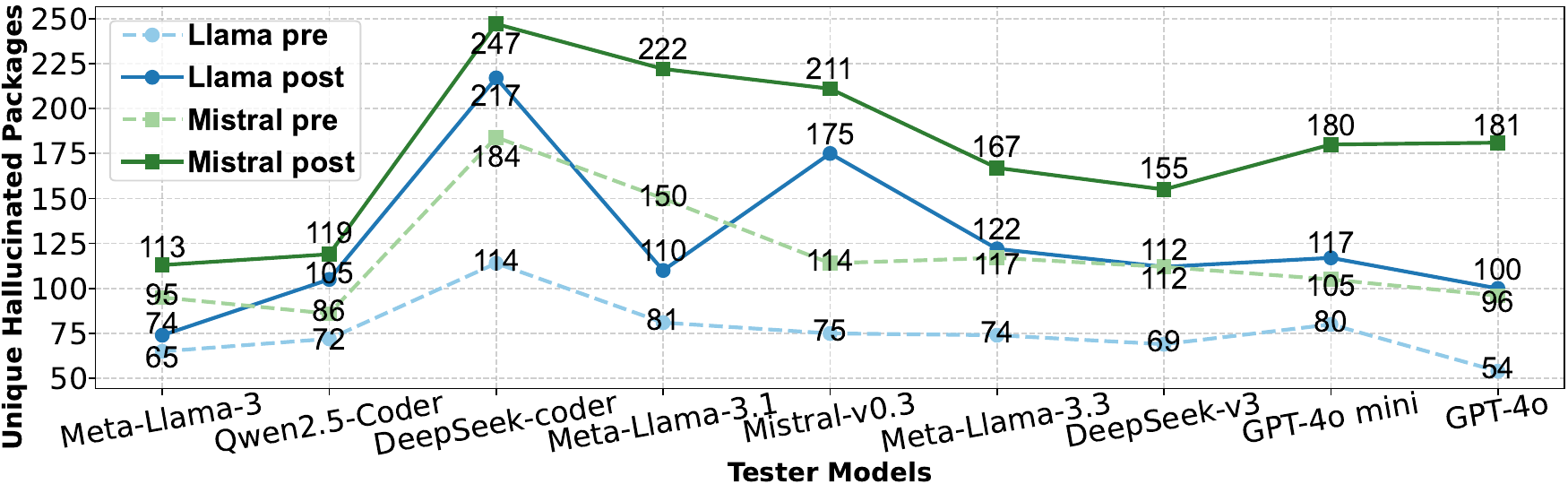}}
\vspace{-0.3cm}
\caption{\fix{Results with Different {\toolname} Input}}
\label{discuss_1}
\vspace{-0.6cm}
\end{figure}

Inspired by Latendresse et al.~\cite{latendresse2024chatgpt}, we examine hallucinated packages to identify placeholders.
\fix{Following the definition of Latendresse et al.~\cite{latendresse2024chatgpt}, two authors independently review all hallucinated packages and resolve disagreements through discussion.}
We find five placeholders: \code{ammonia_library}, \code{some_ml_library}, \code{some_multimedia_sdk}, \code{some-security-sdk}, and \code{some-cloud-language-api}.
For the first three placeholders, the model explicitly includes comments in the generated code to indicate their placeholder status.
\fix{To further analyze the impact of such comments, we ask two authors to independently review the code snippets containing comments.
Their findings are consistent: all hallucinated packages with comments are marked as placeholders or assumptions.
In total, 12 hallucinated packages contain similar comments, including both placeholders and assumptions (e.g., ``Assuming the library is named this'').}
However, when queried for installation commands, the model ignores these comments and still returns the corresponding installation commands, which is the input-conflicting hallucination.
In Figure~\ref{ExamplesA}~(C), the model generates code that includes a placeholder \code{some_ml_library} and a comment clarifying its placeholder state.
Despite this, the model still returns an installation command containing this placeholder.
Similarly, in Figure~\ref{ExamplesA}~(D), the model adds a comment assuming the existence of \code{dsl_parser}.
However, the model ignores the comment when returning an installation command and incorrectly returns \code{pip install dsl_parser}.
These findings indicate that the model may ignore code comments when analyzing the required packages, which motivates further investigation into the influence of code comments on LLM responses in code-related tasks.

\greyboxb{Findings:}{\fix{~(1) For GPT-4o, Package hallucinations occur not only during code generation, but also when assisting with environment configuration, even if the correct code has been provided.} (2) The model may ignore comments when analyzing the required packages in the code.}

\section{Discussion}
\label{sec:discussion}

\subsection{The Effect of Potential Data Contamination}
\fix{To investigate the effect of potential data contamination, according to the ``SourceRank'', we select the first 100 Python packages released after the training cutoff dates of the chosen models (post-cutoff packages).
We run 1000 rounds using all tester models and two target models (Meta-Llama-3.1 and Mistral-v0.3). 
Since the top 100 packages used in Section~\ref{subsec:rq1} are released before the cutoff dates (pre-cutoff packages), we compare the post-cutoff results with those in Section~\ref{subsec:rq1}.}

\fix{The results are shown in Figure~\ref{discuss_1}.
{\toolname} finds more hallucinated packages when using post-cutoff packages.
Compared with pre-cutoff packages, post-cutoff ones yield slightly higher diversity, with the Diversity Index increasing by 0.3\%, 7.6\%, and 24.8\% under DBSCAN parameters $\varepsilon = 0.1, 0.2,$ and $0.3$, respectively. 
Further analysis reveals that tasks derived from these packages contain elements unfamiliar to the models (e.g., model context protocol), and the seed pool includes more phrases. 
When the information sought extends beyond the model’s training data, LLM may fail to provide accurate answers~\cite{gao2023retrieval}, resulting in more hallucinated packages.
Moreover, information outside the training data may introduce greater randomness in responses, thus affecting the diversity of the generated tasks.
Considering that LLMs are rarely applied to tasks involving unknown content, in Section~\ref{subsec:rq1}, we use pre-cutoff packages to evaluate {\toolname}.}

\begin{figure}[t]
\centerline{\includegraphics[width=0.99\linewidth]{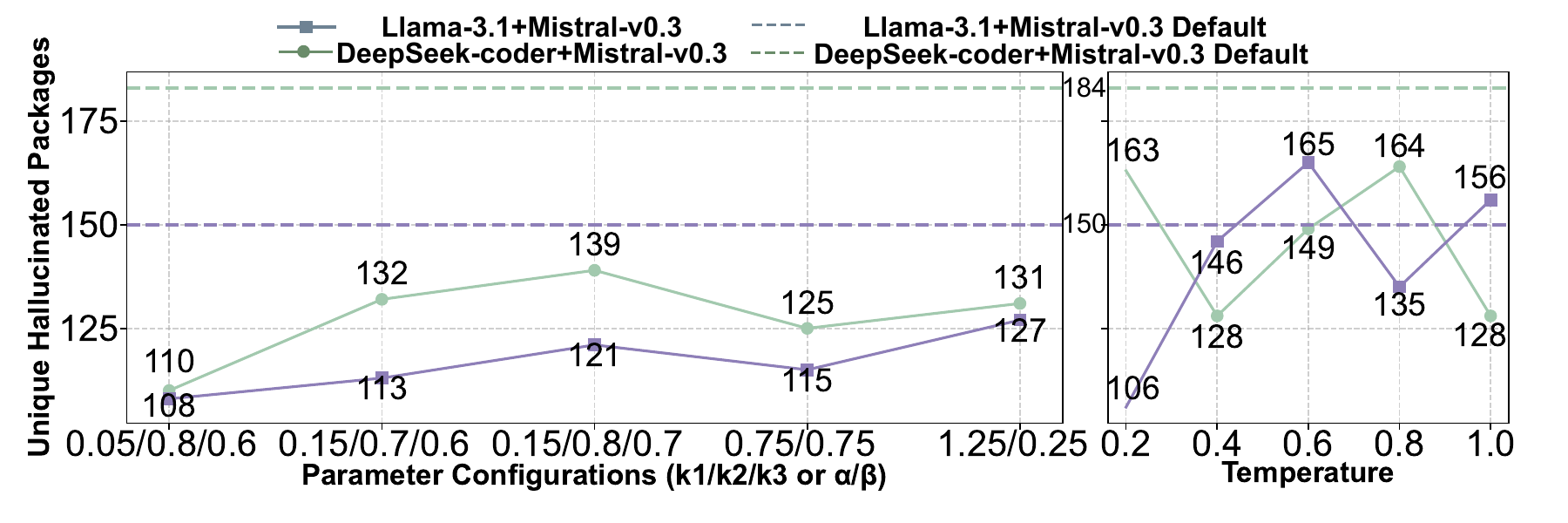}}
\vspace{-0.3cm}
\caption{\fix{The results of different parameters.}}
\label{discuss_2}
\vspace{-0.5cm}
\end{figure}

\subsection{The Effect of Parameter Settings}
\label{sec:Parameter}
\fix{To investigate the impact of {\toolname}'s parameters, we run {\toolname} with different parameter settings on the two model combinations that find the largest number of unique hallucinated packages in RQ1 (i.e., DeepSeek-coder+Mistral-v0.3 and Meta-Llama-3.1+Mistral-v0.3).
For score parameters ($\alpha / \beta$), we test two variants to adjust the impact of ``nonexistentPackage'': 1.25/0.25 and 0.75/0.75.
For power parameters ($k_{1}/k_{2}/k_{3}$), we test three variants: 0.05/0.8/0.6 (reduces the impact of discovering new hallucinated packages); 0.15/0.7/0.6 (reduces the impact of recommending packages); and 0.15/0.8/0.7 (reduces the impact of selection).
Additionally, we test five temperature variants to analyze their impacts.
Other settings are consistent with RQ1.
The Default is the parameter setting used in Section~\ref{evaluation}.}

\fix{Figure~\ref{discuss_2} shows that, for the number of unique hallucinated packages, the parameter variants find fewer hallucinated packages than the Default.
In contrast, the effect of temperature is not significant, with no clear trend observed.
The parameter variants generate less diverse tasks, with Diversity Indices between 85\% and 96\% of the Default, averaged over all clustering parameters.
Temperature also influences diversity: at 0.2, the Diversity Index decreases to 89\%, whereas at 0.4–1.0, it increases slightly (by 1.02\%–1.26\%).
Overall, parameter and temperature settings affect the performance of {\toolname}, but their impact remains limited, and the results consistently outperform those of {\baseline}.}

\subsection{The Impact of Package Hallucinations}
To further assess the impact of package hallucinations on real-world developers, we manually investigate instances of GPT-4o's hallucinated package usage on GitHub~\cite{github} and find that some hallucination packages have been used in the repository.
For instance, we identify that the hallucinated package ``data-analysis-toolkit'' appears in several open-source repositories, such as research-manual\footnote{https://github.com/iHuman-Lab/research-manual}.
Similarly, Lanyado conducts a case study on hallucinated packages~\cite{packagehallucinations2024}, finding that a hallucinated package is downloaded over 30,000 more times than a randomly named package within three months, and is even incorporated into repositories of some commercial companies.
% Similarly, Lanyado conducts a case study of hallucinated packages~\cite{packagehallucinations2024}
% , in which he registers a hallucinated empty package and a randomly named package in the package repository and monitors their downloads. 
% Within three months, this hallucinated package is downloaded over 30,000 more times than the randomly named package and is even used in the repositories of some commercial companies.
% They find that the hallucinated package is downloaded over 30,000 more times than the randomly named package and is even used in the repositories of some commercial companies.
% These findings illustrate that developers are susceptible to package hallucinations during development, which may expose them to malicious attacks.
% Furthermore, researchers and practitioners have developed various LLM agents (such as Devin~\cite{Devin}, Manus~\cite{manus}) that are capable of end-to-end software development and can automatically deploy applications.
% These agents are likely to use hallucinated packages in practice, thereby increasing the risk of malicious attacks.
These findings illustrate that developers can be inadvertently exposed to hallucinated packages during development, potentially increasing their vulnerability to malicious attacks. Moreover, recent advances in LLM-based software development agents, such as Devin~\cite{Devin}, which support end-to-end application development and automatic deployment, further exacerbate this risk, as these agents may unknowingly install hallucinated packages in practice.

\subsection{Potential downstream tasks}
\noindent \textbf{Package Hallucination Mitigation:}
In existing studies, Spracklen et al.~\cite{spracklen2024we} have proposed their insights to mitigate package hallucination.
Spracklen et al.~\cite{spracklen2024we} make a preliminary attempt to mitigate package hallucination using several popular techniques.
Their experimental results show that Retrieval Augmented Generation (RAG), Self-Detected Feedback, and Fine-tuning all help reduce the model PHR, with Fine-tuning proving to be the most effective.
% They filter their dataset to retain only non-hallucinated responses and fine-tune the model.
On the DeepSeek Coder 6B, PHR is reduced by 83\% through fine-tuning.
% Although there is some overlap between the dataset used for fine-tuning and the evaluation dataset, the significant improvement still demonstrates the effectiveness of Fine-tuning.
{\toolname} can provide sufficient data for fine-tuning, enabling more effective mitigation and future evaluation.

\noindent  \textbf{Model Performance Improvement:}
According to the research of Krishna et al.~\cite{krishna2025importing}, PHR is negatively correlated with the performance of the model on the code benchmark.
Therefore, reducing inherent model package hallucinations through techniques such as fine-tuning and model editing is a promising direction for improving model performance in the field of code in the future.
Our work provides sufficient data support and evaluation methods for such follow-up studies.

% Additionally, Krishna et al.~\cite{krishna2025importing} propose mitigating package hallucinations in LLM-based coding tools by checking whether a package is registered before the model's knowledge cutoff date.
% Considering the vast scale of package repositories, such checks could introduce unnecessary overhead, reduce the responsiveness of the tool, and limit the use of auxiliary model methods such as RAG and internet search.
% Moreover, according to the research of Krishna et al.~\cite{krishna2025importing}, PHR is negatively correlated with the performance of the model on the code benchmark.
% Therefore, we believe that it is more valuable to further explore package hallucinations in the model and to achieve package hallucination mitigation for the model itself.
% We plan to make further attempts to mitigate the package hallucination in future work.

% In this paper, we focus on package hallucination testing and do not analyze the code generated by the model.
% However, {\toolname} can generate logical coding tasks, which also gives it the potential for code hallucination testing. 
% We plan to explore this further by integrating existing code hallucination classification methods.

\subsection{\fix{Testing LLM VS. Detecting LLM Hallucinations}}
\fix{Existing studies~\cite{chen2024inside,zhang2023enhancing,quevedo2024detecting,yang2025hallucination,tian2024codehalu} are proposed to detect hallucinations.
Other studies~\cite{tanzil2024chatgpt, liu2023your, prenner2022can,
fan2023automated,liu2024refining,xia2024automated,pan2024lost} can also detect package hallucinations as a side effect.
In contrast, {\toolname} is a testing approach to test LLMs, representing a distinct yet complementary research direction.
The core differences are as follows:}

\noindent  \textbf{\fix{Technique Difference:}} \fix{Whereas existing studies analyze LLM outputs, {\toolname} generates diverse and logical inputs.}

\noindent  \textbf{\fix{Goal Difference:}} \fix{The goal of related studies is to detect whether LLm outputs are hallucinations, while the goal of {\toolname} is to test LLMs.}

\subsection{Threats to Validity}

\noindent \textbf{Limited of Language.}
In this paper, we primarily evaluate {\toolname} on Python.
% Note that {\toolname} is language-independent.
However, the framework is designed to be language-agnostic.
We focus on Python because it is widely used and frequently studied in related work~\cite{krishna2025importing,spracklen2024we}.
In future work, we plan to extend our evaluation to multiple programming languages.
% Actually, its design is decoupled from any specific language, and we discuss its performance in Python because Python is a popular language and has been frequently mentioned in related studies~\cite{krishna2025importing,spracklen2024we}.
% We plan to conduct further evaluations in future work on multiple programming languages.
% 在这篇论文中，我们主要在python语言上评估了{\toolname}。
% However, it is important to note that {\toolname}并不受语言限制。
% 它的设计并不与语言耦合，我们在这讨论python上的效果是因为python是一个非常流程的语言，并在相关研究中都被提及。
% 我们计划在后续的工作中展开多种语言上进行进一步的评估。

\noindent \textbf{Limited Accuracy of Regular Extraction.}
We rely on regular expressions to extract information from the model's output.
Due to the inherent instability of the LLM’s output, the extracted content may not always align with expectations.
To improve the stability of the LLM’s output format, we employ one-shot prompting to guide the output format and implement validation checks on the extracted results.

\noindent \textbf{Limited LLM Sampling Strategy.}
\fix{The LLM sampling strategy may affect the performance of LLM-based methods.
Investigating the impact of sampling strategies on {\toolname} helps to assess its robustness.
Therefore, we examine the effect of the temperature of the tester model in Section~\ref{sec:Parameter}.
For cost considerations, the study is conducted on two model combinations.
We plan to further extend it in future work.}
% We rely on regular expressions to extract relevant information from the model's output. 
% Due to the inherent instability of the LLM's output, we cannot guarantee that the content extracted always matches our expectations.
% In the implementation, to improve the stability of the LLM's output format as much as possible, we prompt the LLM's output format using one-shot and perform checks on the extracted results.

\section{Related Work}
%Here, we discuss state-of-the-art studies in related fields.

\subsection{Fuzzing}
Fuzzing is one of the most popular testing techniques, which can find weaknesses in a program~\cite{zhao2024systematic, liang2018fuzzing}.
Lee et al.~\cite{lee2023learning} group seeds by syntax and semantic similarity and use a customized Thompson sampling method to choose effective mutation strategies for each group.
Liu et al.~\cite{liu2024fuzzinmem} propose mutating inputs directly in memory and using print functions to regenerate files, improving fuzzing for complex file formats.
Yang et al.~\cite{yang2023fuzzing} implement a fully automated API-level fuzzer for automatic differentiation in deep learning libraries.
Hough et al.~\cite{hough2024crossover} exploit dynamic execution information to identify and exchange similar parts of parameter sequences for parametric fuzzing.
Wang et al.~\cite{wang2024tacoma} use fine-grained semantic alignment techniques to generate semantically correct test inputs for fuzzing browsers.
With the widespread use of LLMs, some researchers have attempted to combine LLMs with fuzzing.
Xia et al.~\cite{xia2024fuzz4all} leverage an LLM to generate and mutate inputs for systems that take programming languages or formal languages as inputs.
Eom et al.~\cite{eom2024covrl} combine coverage feedback with an LLM-based mutator using reinforcement learning to test JavaScript engines.
Deng et al.~\cite{deng2024large} propose FuzzGPT, which uses LLMs to generate anomalous programs for fuzzing deep learning libraries.
In contrast to these works, {\toolname} treats the LLM as the target of testing.

\subsection{LLM Jailbreak}
LLM jailbreak refers to exploiting carefully crafted prompts to elicit content that violates service guidelines~\cite{yao2024fuzzllm}.
To better guide defense strategies, existing studies have conducted extensive testing on LLM jailbreak vulnerabilities through red team attacks~\cite{yao2024fuzzllm,yu2023gptfuzzer,deng2023masterkey}.
Yao et al.~\cite{yao2024fuzzllm} use templates to preserve prompt structure and isolate jailbreak features as constraints, creating an automated framework to test and find jailbreak vulnerabilities in LLMs.
Yu et al.~\cite{yu2023gptfuzzer} exploit manually crafted templates as initial input and mutate them to generate new templates.
Additionally, some studies have also made efforts to defend against LLM jailbreak.
Zhang et al.~\cite{zhang2025jbshield} adjust the target LLM's hidden representations by enhancing toxic concepts and weakening jailbreak concepts, ensuring the LLM generates safe content.
Wang et al.~\cite{wang2024defending} leverage reverse inference on the initial responses of the LLM  to reveale the actual intent behind the original prompt. They then re-prompt the LLM to generate responses in a way that mitigates potential jailbreak attacks.
Extensive studies have been conducted through red team attacks to test LLMs for jailbreak vulnerabilities and guide defense strategies.
However, studies on hallucinations are relatively scarce.
{\toolname} fills this gap, laying the foundation for better mitigation of LLM package hallucinations, which is a high-risk hallucination type.

\subsection{\fix{LLM Hallucination}}
\fix{Despite recent progress, LLMs still generate hallucinations~\cite{krishna2025importing}, which can be categorized into input-, context-, and fact-conflicting types~\cite{zhang2023siren}.
To reduce the impact of hallucinations, researchers conduct studies on detecting and mitigating hallucinations.}
Jones et al.~\cite{jones2023teaching} utilize prefix-tuning on synthetic tasks to optimize the system message, then transfer this message to realistic tasks to reduce hallucination.
Chen et al.~\cite{chen2024inside} propose to explore the dense semantic information retained within LLMs' internal states for hallucination detection.
Zhang et al.~\cite{zhang2023enhancing} introduce a reference-free, uncertainty-based method to detect hallucinations in LLMs.
Quevedo et al.\cite{quevedo2024detecting} construct two simple classifiers for hallucination detection using four numerical features, using supervised learning.
Yang et al.~\cite{yang2025hallucination} use metamorphic testing to mutate the model response and evaluate whether the mutated content is correct through LLM to detect hallucinations.
Note that their mutation method is still based on the assumption of a unique answer.

\fix{Given the application of LLMs in code-related tasks, researchers study hallucinations in the code domain~\cite{yu2024fight}.
% They find that popular LLMs produce code hallucinations 
Liu et al.~\cite{liu2024exploring} explore hallucinations in code generation and propose a novel classification for code hallucinations.}
Jain et al.~\cite{jain2024mitigating} mitigate API hallucinations in low-frequency APIs through documentation augmented generation.
Tian et al.~\cite{tian2024codehalu} propose a hallucination detection algorithm based on execution validation and a code hallucination classification method.
\fix{Spracklen et al.~\cite{spracklen2024we} and Krishna et al.~\cite{krishna2025importing} further investigate a special type of code hallucinations, i.e.,  package hallucination.
They define package hallucination as an LLM generates code that either recommends or contains a reference to a package that is not registered in the appropriate package repository or is first registered after the model’s knowledge cutoff date.
These hallucinations pose security risks, as attackers may register phantom packages with malicious code, which LLMs then recommend to developers~\cite{spracklen2024we,krishna2025importing,packagehallucinations2023,packagehallucinations2024}.
Considering the security risks posed by package hallucination, we propose {\toolname} and apply it to package hallucination.
Different from existing methods for hallucination detection and mitigation, {\toolname} focuses on generating tasks to trigger model hallucinations, which is a method similar to red team attacks.}

\section{Conclusion and Future Work}
In this paper, we present {\toolname}, a novel phrase-based fuzzing framework to test LLMs for package hallucinations.
By automatically generating diverse coding tasks based on phrases, {\toolname} extensively tests LLMs for package hallucinations.
Through extensive evaluation, we demonstrate that {\toolname} can trigger package hallucination across all selected models, \fix{find 2.60x more unique hallucinated packages compared with {\baseline}, and generate more diverse tasks.}
We further test the model GPT-4o and find 46 unique hallucinated packages.
\fix{Our analysis shows that for GPT-4o, package hallucinations not only occur in code generation but also occur when assisting with environment configuration.}
% This highlights {\toolname}'s potential to assist researchers in enhancing the security of LLMs.

In the future, we plan to conduct more extensive testing on more LLMs and more types of languages, and promote the study of LLM package hallucination mitigation through open-sourcing results.
Additionally, we hope to further expand our framework on code hallucination.
% We also consider adjusting the parameters used in our framework through repeated experiments.

\label{sec:conclusion}

\section{Acknowledgments}
This research is supported by the National Key R\&D Program of China (No. 2024YFB4506400).

% \newpage
\balance
\bibliographystyle{IEEEtran}
\bibliography{references}
% \bibliographystyle{ACM-Reference-Format}
% \bibliography{references}

\end{document}